\documentclass[12pt]{article} 
\def\cdate{{August 9, 2014}}
\usepackage{amsfonts,amsmath,latexsym,amssymb,txfonts} 
\usepackage[american]{babel}

\makeatletter
\def\timenow{%
\@tempcnta=\time \divide\@tempcnta by 60 \number\@tempcnta:\multiply
\@tempcnta by 60 \@tempcntb=\time \advance\@tempcntb by -\@tempcnta
\ifnum\@tempcntb <10 0\number\@tempcntb\else\number\@tempcntb\fi}

\newcounter{outputpage}

\renewcommand{\@evenhead}{\raisebox{0pt}[\headheight][0pt]{\vbox{\hbox
to \textwidth{\thepage\hfil\strut\textit{\leftmark}}
}}}
\renewcommand{\@oddhead}{\raisebox{0pt}[\headheight][0pt]{\vbox{\hbox
to \textwidth{\textit{\rightmark}\hfil\strut\thepage}
}}}

\renewcommand{\@oddfoot}
{\vbox{
\vspace{3pt}
\hfil
{\scriptsize\textit{
\stepcounter{outputpage}
\hfill\hfill
}}
\hfil
}}

\renewcommand{\@evenfoot}
{\vbox{
\vspace{3pt}
\hfil
{\scriptsize\textit{
\stepcounter{outputpage}
\hfill\hfill
}}
\hfil
}}

\makeatother

\def\II{{\mathbb I}} 
\def\RR{{\mathbb R}}

\def\a{\alpha} 
\def\b{\beta} 
\def\g{\gamma} 
\def\d{\delta}

\def\m{\mu} 
\def\n{\nu}

\def\s{\sigma}

\def\a{\alpha} 
 
\def\b{\beta} 
\def\g{\gamma} 
\def\d{\delta}

\def\m{\mu} 
 
\def\n{\nu} 
 
\def\s{\sigma}

\def\tr{\mathrm{ tr\,}} 
\def\Tr{\mathrm{ Tr\,}}

\def\vol{\mathrm{ vol\,}}

\def\be{\begin{equation}} 
\def\ee{\end{equation}} 
\def\bea{\begin{eqnarray}} 
\def\eea{\end{eqnarray}} 
\def\bed{\begin{definition}{\ }}
\def\eed{\end{definition}}
\def\bd{\begin{description}}
\def\ed{\end{description}}
\def\bc{\begin{center}}
\def\ec{\end{center}}

\newtheorem{theorem}{Theorem}

\newtheorem{definition}{Definition}

\def\sideremark#1{\ifvmode\leavevmode\fi\vadjust{\vbox to0pt{\vss
 \hbox to 0pt{\hskip\hsize\hskip1em
 \vbox{\hsize2cm\tiny\raggedright\pretolerance10000
 \noindent #1\hfill}\hss}\vbox to8pt{\vfil}\vss}}}%
\begin{document}


\begin{titlepage}
\thispagestyle{empty}
\null
\vspace{-3cm}
\hspace*{50truemm}{\hrulefill}\par\vskip-4truemm\par
\hspace*{50truemm}{\hrulefill}\par\vskip5mm\par
\hspace*{50truemm}{{\large\sc New Mexico Tech {\rm
(\cdate)
}}}
\vskip4mm\par
\hspace*{50truemm}{\hrulefill}\par\vskip-4truemm\par
\hspace*{50truemm}{\hrulefill}
\par
\bigskip
\bigskip
\vfill
\centerline{\huge\bf Heat Determinant}
\bigskip
\centerline{\huge\bf on Manifolds}
\bigskip
\bigskip
\bigskip
\centerline{\Large\bf Ivan G. Avramidi and Benjamin J. Buckman}
\bigskip
\centerline{\it New Mexico Institute of Mining and Technology}
\centerline{\it Socorro, NM 87801, USA}
\centerline{\it E-mail: iavramid@nmt.edu, bbuckman@nmt.edu}
\bigskip
\bigskip
\centerline{\cdate}
\vfill
{\narrower
\par
We introduce and study new invariants associated with Laplace type elliptic partial
differential operators on manifolds. These invariants are constructed by using the
off-diagonal heat kernel; they are not pure spectral invariants, that is, they depend
not only on the eigenvalues but also on the corresponding eigenfunctions in a 
non-trivial way. We compute the first three low-order invariants explicitly.
}


\end{titlepage}

\section{Introduction}
\setcounter{equation}{0}

The heat kernel is one of the most important tools of global analysis, spectral
geometry, differential geometry and mathematical physics, in particular,
quantum field theory \cite{hadamard23,berline92,gilkey95,hurt83}. In quantum field
theory the main objects of interest are described by the Green functions of
self-adjoint elliptic partial differential operators on manifolds and their
spectral invariants such as the functional determinants. In spectral geometry
one is interested in the relation of the spectrum of natural elliptic partial
differential operators to the geometry of the manifold, more precisely, one
studies the question: ``To what extent does the spectrum of a differential
operator determine the geometry of the underlying manifold?''

There are also non-trivial links between the spectral invariants and the
non-linear completely integrable evolution systems, such as Korteweg-de Vries
hierarchy (see, e.g. \cite{hurt83}). In many interesting cases such
systems are, in fact, infinite-dimensional Hamiltonian systems, and the
spectral invariants of a linear elliptic partial differential operator are
nothing but the integrals of motion of this system.

Instead of studying the spectrum of a differential operator directly one
usually studies its spectral functions, that is, spectral traces of some
functions of the operator, such as the zeta function, and the heat trace.
Usually one does not know the spectrum exactly; that is why, it becomes very
important to study various asymptotic regimes. It is well known, for example,
that one can get information about the asymptotic properties of the spectrum by
studying the short time asymptotic expansion of the heat trace. The
coefficients of this expansion, called the heat trace coefficients (or global
heat kernel coefficients), play very important role in spectral geometry and
mathematical physics \cite{hurt83,gilkey95}.

The simplest case of a Laplace operator on a compact manifold without boundary
is well understood and there is a vast literature on this subject, see
\cite{gilkey95} and the references therein. 
For a Laplace type operator on a compact manifold without boundary there is a
well defined local asymptotic expansion of the heat kernel, which enables one
to compute its diagonal and then the heat trace by directly integrating the
heat kernel diagonal; this gives all heat trace coefficients. 
However, many ideas and techniques do not
apply directly in more general cases. 

The existence of non-isometric isospectral manifolds
demonstrates that the {\it spectrum alone does not determine the geometry}
(see, e.g. \cite{berger03}). That is why, we propose to study more general invariants of
partial differential operators that are not spectral invariants, that is, they
depend not only on the eigenvalues but also on the eigenfunctions, and,
therefore, contain much more information about the geometry of the manifold.

In this paper we propose to study {\it new heat invariants} of second-order
Laplace type elliptic partial
differential operators acting on sections of vector bundles over Riemannian
manifolds. Our goal 
is to develop a comprehensive methodology for such invariants in the
same way as the theory of the standard heat trace invariants. Namely,
we will define and study new heat invariants of differential operators
and compute
explicitly some leading terms of the asymptotic expansion of new heat
invariants.

Our main result can be formulated as follows.
\begin{theorem}
Let $(M,g)$ be a smooth compact $n$-dimensional
Riemannian manifold without boundary
with metric $g$
and $\mathcal{V}$ be a vector bundle over $M$ of dimension $N$.
Let $\nabla$ be a connection on the vector bundle and $Q$ be an endomorphism
of the bundle $\mathcal{V}$. Let $\Delta=g^{\mu\nu}\nabla_\mu\nabla_\nu$
be the Laplace operator and 
$L: C^\infty(\mathcal{V})\to C^\infty(\mathcal{V})$
be the Laplace type partial differential operator of the form
$L=-\Delta+Q$. Let $U(t;x,x')$ be the heat kernel of the operator $L$,
\be
P_{\mu\nu'}(t;x,x')=\tr U^*(t;x,x')\nabla_\mu\nabla_{\nu'}U(t;x,x'),
\ee
where $\tr$ is the fiber trace,
and $K(t)$ be the functional defined by
\be
K(t)=\int_{M\times M}dx\,dx'\; \det P_{\mu\nu'}(t;x,x')\,.
\ee
Then there is an asymptotic expansion as $t\to 0$
\be
K(t)\sim (4\pi)^{-n^2}\left(\frac{\pi}{2n}\right)^{n/2}t^{-n\left(n+\frac{1}{2}\right)}
\sum_{k=0}^\infty t^k B_k,
\ee
where 
\be
B_k=\int_M dv\; b_k,
\ee
$dv$ is the Riemannian volume element
and $b_k$ are differential polynomials in the Riemann
curvature, the bundle curvature and the endomorphism $Q$
with some universal numerical coefficients that depend 
only on the dimensions $n$ and $N$.
The low order coefficients are
\bea
b_0 &=& \frac{1}{2}N^n,
\\
b_1 &=& N^{n}\frac{12n^2-n+10}{72n}R-nN^{n-1}\tr Q,
\eea
\bea
b_2 &=&
N^n\Biggl\{\frac{20n^4-8n^3-11n^2-6n+6}{144n^2} R^2
+ \frac{4n^3+11n^2+n-4}{120n^2} \nabla_\m \nabla^\m R
\nonumber\\
&&
+ \frac{-24n^3+84n^2-576n+385}{4320n^2} R_{\m\n}R^{\m\n}
+ \frac{8n^3-8n^2-18n+15}{1440n^2} R_{\a\b\m\n}R^{\a\b\m\n}
\Biggr\}
\nonumber\\
&&
+ \frac{n^3-n^2+3n-12}{12n^2} N^{n-1} \tr {\cal R}_{\m\n} {\cal R}^{\m\n}
+ \frac{-12n^3-4n^2+n+2}{12n} N^{n-1} R \tr Q
\nonumber\\
&&
- \frac{n+2}{6} N^{n-1} \tr \nabla^\m \nabla_\m Q
+ n N^{n-1} \tr Q^2
+ n(n-1) N^{n-2} ( \tr Q )^2 \,.
\eea
Here $R_{\mu\nu\alpha\beta}$, $R_{\mu\nu}$ and $R$ are the
Riemann tensor, the Ricci tensor and the scalar curvature
respectively, and $\mathcal{R}_{\mu\nu}$ is the curvature
of the bundle connection.
\end{theorem}
Of course, the derivative terms can be neglected on manifolds without boundary.

This paper is organized as follows. In Sec. 2 we describe the necessary geometric 
framework and define the heat kernel and some invariants, such as the heat trace and the
heat content. In Sec. 3. we define a new invariant called the heat determinant for scalar
operators and show that on manifolds without boundary it is trivial.
In Sec. 4 we define an alternative invariant that we call the heat determinant on vector bundles.
In Sec. 5 we introduce the machinery of standard off-diagonal heat kernel asymptotics
and compute the heat content asymptotics.
In Sec. 6 we compute the determinant of the mixed derivatives of the heat kernel.
In Sec. 7 we establish the asymptotics of the heat determinant and in Sec. 8 we compute
some low-order coefficients of this expansion. Some of the technical formulas for the
derivatives of the Synge function and the parallel transport operator are listed in the Appendix.


\section{Heat Kernel}
\setcounter{equation}{0}

Let $(M,g)$ be a compact Riemannian manifold of dimension $n$
and ${\cal V}$ be a vector bundle over the manifold $M$
with a typical fiber $V$ of dimension $N$. 
First of all, we fix notation.
We denote the local coordinates in a chart by
$x=(x^\mu)$, $\mu=1,\dots,n$, 
and use the Einstein summation convention.
The components of the metric
in the coordinate basis are denoted by $g_{\mu\nu}$,
the determinant of the metric is denoted by
$g=\det g_{\mu\nu}$,
the Levi-Civita symbol is denoted by
$\varepsilon_{\mu_1\dots\mu_n}$,
the covariant Levi-Civita symbols are denoted by
$E_{\mu_1\dots\mu_n}=g^{1/2}\varepsilon_{\mu_1\dots\mu_n}$
and $E^{\mu_1\dots\mu_n}=g^{-1/2}\varepsilon^{\mu_1\dots\mu_n}$.
Let $dx=dx^1\cdots dx^n$ be the Lebesgue measure in a local chart,
and $dv=dx\, g^{1/2}$, 
be the Riemannian volume element;
further, let $\left<\varphi,\psi\right>=\tr \varphi^*\otimes \psi$ be
the fiber inner product, where $\varphi^*$ is the dual section
and $\tr$ is the fiber trace.
We use parenthesis to denote symmetrization over
all indices included and square brackets to denote the 
anti-symmetrization. The indices excluded from (anti)-symmetrization
are separated by vertical lines.

We will consider two-point functions and tensors such as
$G(x,x')$. Then the derivatives of such two functions with
respect to $x^\mu$ will be denoted by $\nabla_\mu G$ 
(or by a semicolon $G_{;\mu}$)
and
the derivatives with respect to $x'^{\nu}$ will be denoted by
$\nabla_{\nu'}G$ (or by a semicolon $G_{;\nu'}$). 
For convenience we introduce special notation for the coincidence limit
of two-point functions; we will denote it by square brackets
\be
[f]=\lim_{x\to x'}f(x,x')\,.
\ee

In case when there is a boundary $\partial M$, we assume that
the boundary is smooth so that the inward pointing unit normal 
$N$ is well defined. We use the natural orientation of the boundary
and denote the local coordinates on the boundary by 
$\hat x=(\hat x^i)$, $i=1,\dots, n-1$, so that the local equation
of the boundary is $x^\mu=x^\mu(\hat x)$. The Lebesgue measure
on the boundary is denoted by $d\hat x=d\hat x^1\dots d\hat x^{n-1}$.
The induced Riemannian metric on the boundary is denoted by
$\hat g_{ij}=\frac{\partial x^\mu}{\partial \hat x^i}
\frac{\partial x^\nu}{\partial \hat x^j}g_{\mu\nu}$ 
and the Riemannian volume element on the boundary
by $d\hat v=d\hat x\,\hat g^{1/2}$, where 
$\hat g=\det \hat g_{ij}$.

Let $C^\infty({\cal V})$ be the space of smooth sections
of the bundle ${\cal V}$. We define a natural invariant $L^2$
inner product on $C^\infty({\cal V})$ by 
$(\varphi,\psi)=\int_M dv \left<\varphi,\psi\right>$
and the corresponding norm $||\varphi||=\sqrt{(\varphi,\varphi)}$.
The completion of the space $C^\infty({\cal V})$ in this norm
defines the Hilbert space $L^2({\cal V})$.
The $L^2$ trace of a trace-class operator $G$
will be denoted by $\Tr$, that is,
\be
\Tr G=\int\limits_M dv\,\tr G(x,x)\,,
\ee
where $\tr$ is the fiber trace and $G(x,x')$ is the integral kernel of the operator $G$.

The components of a connection one form
on the vector bundle ${\cal V}$ are denoted by ${\cal A}_\mu$.
We introduce the covariant exterior derivative of sections of ${\cal V}$
by
\be
D\varphi=d\varphi+{\cal A}\wedge \varphi\,.
\ee
Then, obviously,
\be
D^2\varphi={\cal R}\varphi\,,
\ee
where
\be
{\cal R}=d{\cal A}+{\cal A}\wedge{\cal A}\,
\ee
is the curvature of the connection ${\cal A}$.

Let $L: C^\infty({\cal V})\to C^\infty({\cal V})$ be a second-order
formally self-adjoint elliptic partial differential operator with a positive
definite leading symbol. If the leading symbol of the operator $L$ is scalar
then the operator $L$ is called of {Laplace type}; in this case one can
define a Riemannian metric $g_{\mu\nu}$
by the leading symbol of the operator $L$, a 
connection $\nabla$ on the vector bundle ${\cal V}$ and a
self-adjoint
endomorphism
$Q$ of the vector bundle ${\cal V}$ so that
\be
L=-\Delta+Q,
\label{25xx}
\ee
where $\Delta=g^{\mu\nu}\nabla_\mu\nabla_\nu$ is the Laplacian.
In the case when there is a (smooth) boundary $\partial M$
we assume that some suitable boundary
conditions are imposed, either Dirichlet
\be
\varphi\Big|_{\partial M}=0
\ee
or Neumann 
\be
\nabla_N\varphi\Big|_{\partial M}=0.
\ee
Here, $\nabla_N=N^\mu\nabla_\mu$ is the normal derivative
and $N^\mu$ is the inward pointing unit normal 
to the boundary.

It is well known \cite{gilkey95} that if the manifold $M$ is compact then
the operator $L$ has only point spectrum consisting of real discrete
eigenvalues with finite multiplicities 
bounded from below
\be
\lambda_1\le \lambda_2\le\lambda_3\le \cdots,
\ee
where each eigenvalue is counted with its (finite) multiplicity.
The spectrum $(\lambda_k,\varphi_k)_{k=1}^\infty$ of the operator $L$ is determined by
\be
L\varphi_k=\lambda_k \varphi_k\,;
\ee
without loss of generality we can take the sequence of the 
eigensections to be orthonormal, that is,
\be
(\varphi_k,\varphi_m)
=\delta_{km}\,.
\ee

Then for $t>0$ the heat semigroup $\exp(-tL)$ is a bounded
operator with the integral kernel
\bea
U(t;x,x') &=&
\sum_{k=1}^\infty e^{-t\lambda_k}\varphi_k(x)\otimes\varphi^*_k(x'),
\eea
called the  heat kernel of the operator $L$.
The heat kernel satisfies the heat equation
\be
(\partial_t+L)U(t;x,x')=0
\ee
with the initial condition
\be
U(0;x,x')=\delta(x,x'),
\ee
where we use the notation 
$
\delta(x,x')=g^{-1/4}(x)g^{-1/4}(x')\delta(x-x')
$
for  the covariant delta-function.

We would like to define some invariants of the operator $L$, 
that we call 
heat invariants, constructed entirely from the heat 
kernel of the operator $L$
without any additional ingredients.
There are two type of invariants. The usual ones, called spectral invariants,
depend only on the eigenvalues and do not depend on the eigensections.
More general heat invariants depend on both the eigenvalues and the eigensections.

One of the best known invariants is the {\it heat trace} 
which is obtained by integrating of the heat kernel diagonal
\be
\Theta(t)=
\Tr\exp(-tL)=
\int\limits_{M}dv\,\tr U(t;x,x)
=\sum_{k=1}^\infty
e^{-t\lambda_k}\,.
\ee
This is obviously a spectral invariant of the operator $L$ since it only
depends on the eigenvalues of the operator but does not depend on the
eigenfunctions.

For a scalar operator acting on smooth real functions 
one can define another invariant (called the {\it heat content})
by integrating the off-diagonal heat
\bea
\Pi(t) &=& \int\limits_{M\times M} dv\,dv'\;
U(t;x,x')=
\sum_{k=1}^\infty e^{-t\lambda_k}\left|\Phi_k\right|^2,
\eea
where
\be
\Phi_k=\int\limits_{M}dv\,
\varphi_k\,.
\ee

For general operators acting on sections of a vector bundle this
obviously does not work since the off-diagonal heat kernel is not 
a scalar function and the trace of the off-diagonal heat kernel
is not invariant. One could, of course, define an invariant with the help
of a section $\psi$ of the bundle $\mathcal{V}$ by
\be
\tilde \Pi(t;\psi)=\int_{M\times M}dv\,dv'\; \left<\psi, U(t;x,x')\psi(x')\right>,
\ee
with
$\left<\cdot,\cdot\right>$ the fiber inner product, however, this
introduces an additional ingredient, namely, an additional section,
which is not an invariant of the operator only.
What we try to do in this paper is rather different, we want to define 
an invariant entirely in terms of the heat kernel without the need for any
additional ingredients.

\section{Heat Determinant of Scalar Operators}
\setcounter{equation}{0}


Let us first consider a general scalar Laplace type operator
(\ref{25xx}) acting on real smooth functions.
We propose to study a new invariant defined as follows.
First, we introduce the tensor of mixed derivatives
of the heat kernel
\be
\tilde P_{\mu\nu'}(t;x,x') = \nabla_\mu\nabla_{\nu'}U(t;x,x')\,.
\ee
Its spectral representation 
takes the form
\be
\tilde P_{\mu\nu'}(t;x,x')=\sum_{k=1}^\infty e^{-t\lambda_k}
\nabla_\mu\varphi_{k}(x)\nabla_{\nu'}\varphi_{k}(x')\,.
\ee
Now, we notice that $\det \tilde P_{\mu\nu'}$ is a bi-scalar density of weight
$1$.
Therefore, we can define a new invariant 
(that we call the {\it heat determinant})
by integrating the determinant with the non-invariant
Lebesgue measure $dx$ instead of the invariant Riemannian
measure $dv=dx\,g^{1/2}$
\bea
\tilde K(t) &=& \int\limits_{M\times M}dx\;dx'\; 
\det \tilde P_{\mu\nu'}(t;x,x')\,.
\eea

Further, we define some new invariants that measure the correlations
between eigenfunctions,
\bea
\tilde \Phi_{k_1\dots k_n} &=&
\int\limits_M d\varphi_{k_1}\wedge\cdots\wedge d\varphi_{k_n}
\,.
\eea
By using the equation
\be
d\varphi_{k_1}\wedge\cdots\wedge d\varphi_{k_n}
=d\left(\varphi_{k_1}\,d\varphi_{k_2}\wedge\cdots\wedge d\varphi_{k_n}\right),
\ee
and the Stokes theorem it is easy to see that 
\bea
\tilde \Phi_{k_1\dots k_n}
& =& \int\limits_{\partial M} \varphi_{k_1}\,
d\varphi_{k_2}\wedge\cdots\wedge d\varphi_{k_n}.
\eea
Thus, for manifolds without boundary or with Dirichlet boundary conditions
all invariants $\tilde \Phi_{k_1\dots k_n}$ vanish,
\be
\tilde \Phi_{k_1\dots k_n}=0\,.
\ee
Also, it is obvious that the invariants $\tilde \Phi_{k_1\dots k_n}$ are completely
antisymmetric in all their indices
$k_1, \dots, k_n$ and, therefore, vanish
if any of the indices are equal.

Now, by using the definition of the heat determinant one can express 
the heat determinant in terms of the spectral form
\be
\tilde K(t)=\frac{1}{n!}
\sum_{k_1,\dots , k_n=1}^\infty
\exp\left\{-t(\lambda_{k_1}+\cdots+\lambda_{k_n})\right\}
\left|\tilde \Phi_{k_1\dots k_n}\right|^2 \,,
\ee
Further, by using the antisymmetry of the invariants $\tilde \Phi_{k_1\dots k_n}$
we can reorder the indices in the increasing order, which leads to 
a combinatorial factor $n!$; thus,
\be
\tilde K(t)=\sum_{1\le k_1<k_2<\cdots< k_n}
\exp\left\{-t(\lambda_{k_1}+\cdots+\lambda_{k_n})\right\}
\left|\tilde \Phi_{k_1\dots k_n}\right|^2\,.
\ee

One can show that for the matrix $\tilde P=(\tilde P_{\mu\nu'})$
of mixed derivatives $\tilde P_{\mu\nu'}=\nabla_\mu\nabla_{\nu'}U$ 
of any scalar two-point function
$U$ there holds
\bea
\det \tilde P &=& 
\frac{1}{n}\partial_{\mu}\partial_{\nu'}\left(U(\det \tilde P)\tilde P^{-1}{}^{\nu'\mu}\right)\,,
\eea
where $\tilde P^{-1}{}^{\nu'\mu}$ is the inverse matrix.
Indeed, we have
\bea
\det \tilde P 
&=&
\frac{1}{n}\partial_{\mu}\partial_{\nu'}G^{\nu'\mu},
\eea
where
\bea
G^{\nu'\mu}&=&
\frac{1}{(n-1)!}
\varepsilon^{\mu\mu_1\dots\mu_{n-1}}\varepsilon^{\nu'\nu'_1\dots\nu'_{n-1}}
\tilde P_{\mu_1\nu'_1}\cdots \tilde P_{\mu_{n-1}\nu'_{n-1}}U\,.
\eea
This tensor is defined for an arbitrary matrix $\tilde P$.
One can easily show that in the case when the matrix
$\tilde P$ is not degenerate this tensor is equal to 
\bea
G^{\nu'\mu}&=&
(\det \tilde P)\tilde P^{-1}{}^{\nu'\mu}U\,.
\eea
Therefore, 
by using the Stokes theorem,
the heat determinant takes the form
\be
\tilde K(t)=\int\limits_{\partial M\times\partial M} d\hat x\, d\hat x'\,
\frac{1}{n}N_\mu N_{\nu'}G^{\nu'\mu}(t;x,x')\,.
\ee
In particular, for manifolds without boundary and for Dirichlet boundary conditions
the heat determinant vanishes
\be
\tilde K(t)=0\,.
\ee

\section{Heat Determinants on Vector Bundles}
\setcounter{equation}{0}

Notice that the definition of the heat determinant above 
does not work for general operators
acting on sections of a vector bundle.
This is because the mixed derivtive
$\nabla_\mu\nabla_{\nu'}U(t;x,x')$ is
a section of the external product bundle
$\mathcal{V}\boxtimes\mathcal{V}^*$
(that is, it is a section of $\mathcal{V}$ at the point $x$
and a dual section at the point $x'$, in addition to being
a covector at both these points).
Therefore, we need to modify it accordingly.
Also, we would like to have an invariant that does not vanish
on manifolds without boundary.
The heat determinant can now be defined as follows.
First,  to define an invariant tensor we multiply the mixed
derivative of the heat kernel by the dual heat kernel
\be
P_{\mu\nu'}(t;x,x')=\tr U^*(t;x,x')\nabla_\mu\nabla_{\nu'}U(t;x,x')\,.
\label{310xx}
\ee
Recall that for a self-adjoint operator $L$
\be
U^*(t;x,x')=U(t;x',x)\,.
\ee
This quantity is the bi-covector at the points $x$ and $x'$, and, therefore,
its determinant is a bi-scalar density. Thus,
in the same way as we defined the heat determinant for the scalar operators
we can define
\bea
K(t) &=& \int\limits_{M\times M}dx\;dx'\,
\det P_{\mu\nu'}(t;x,x')\,.
\eea

Let us define the one-forms $\Phi^k_{l}=\Phi^k_{l\,\mu}dx^\mu$ by
\be
\Phi^k_{l}=\left<\varphi_k, D\varphi_l\right>\,,
\ee
where
\be
\Phi^k_{l\,\mu}=\left<\varphi_k,\nabla_\mu\varphi_l\right>\,,
\ee
and the invariants
\bea
C^{k_1\dots k_n}_{l_1\dots l_n} 
&=&
\int\limits_M \Phi^{k_1}_{l_1}\wedge\cdots\wedge \Phi^{k_n}_{l_n}
\,.
\eea
Then the tensor (\ref{310xx}) takes the form
\be
P_{\mu\nu'}(t;x,x')=\sum_{k,l=1}^\infty e^{-t(\lambda_k+\lambda_l)}
\Phi^k_{l\,\mu}(x)\left(\Phi^{k}_{l\,\nu'}(x')\right)^*\,,
\ee
and the heat determinant can be written in the form
\be
K(t)=\frac{1}{n!}
\sum_{k_1,l_1,\dots , k_n, l_n=1}^\infty
\exp\left\{-t(\lambda_{k_1}+\lambda_{l_1}+\cdots+\lambda_{k_n}+\lambda_{l_n})\right\}
\left|C_{l_1\dots l_n}^{k_1\dots k_n}\right|^2\,.
\ee

The advantage of this invariant is that it is defined entirely in terms of the heat kernel
without the need for any additional ingredients, like a section or the
parallel transport operator or another differential operator.
Further, 
we can define the corresponding zeta function by
the combined Laplace-Mellin transform, that is,
\be
Z(s,\lambda)
=\frac{1}{\Gamma(s)}\int\limits_0^\infty
dt\, t^{s-1} e^{t\lambda}K(t)\,,
\ee
where $\lambda$ has a sufficiently large negative real part and $s$
has a sufficiently large positive real part.
Also, this heat determinant does not vanish for manifolds without boundary
and defines a new invariant of operators on manifolds without boundary.

\section{Heat Kernel Asymptotics}
\setcounter{equation}{0}

We consider manifolds without boundary.
We will extensively use the machinery of two-point geometric
functions such as the Synge function (see, for example
\cite{synge60,dewitt65,avramidi00,avramidi13}).
The Synge function $\sigma(x,x')$ is defined as
one-half the square of the geodesic
distance between the points $x$ and $x'$.
At least for sufficiently 
close points $x$ and $x'$ this function is well
defined and smooth. 

We use the following notation:
each additional index  denotes the covariant derivative with respect
to $x^\mu$ and each primed index denotes the 
covariant derivative with respect
to $x'^\nu$, e.g.  $\sigma_\mu=\nabla_\mu\sigma$,
$\sigma_{\nu'}=\nabla_{\nu'}\sigma$,
$\sigma_{\mu\nu'}=\nabla_\mu\nabla_{\nu'}\sigma$,
$\sigma_{\mu\nu}=\nabla_\mu\nabla_\nu\sigma$,
etc.
Let $\gamma^{\nu\mu'}$ be the matrix inverse to the matrix
$\sigma_{\nu'\mu}$. Recall the identities satisfied by these matrices
\cite{avramidi00}
\be
\sigma=\frac{1}{2}\sigma^\mu\sigma_\mu=\frac{1}{2}\sigma^{\mu'}\sigma_{\mu'}
\ee
\be
\sigma_{\nu'\mu}\sigma^{\mu}=\sigma_{\nu'}, \qquad
\sigma_{\nu'\mu}\sigma^{\nu'}=\sigma_{\mu},
\ee
\be
\gamma^{\mu\nu'}\sigma_{\mu}=\sigma^{\nu'}, \qquad
\gamma^{\mu\nu'}\sigma_{\nu'}=\sigma^{\mu}.
\ee

The determinant of the matrix $\sigma_{\mu\nu'}$ defines the 
Van Fleck-Morette determinant
\cite{dewitt65,avramidi00}
\be
\Delta(x,x')=g^{-1/2}(x)g^{-1/2}(x')
\det\left(-\sigma_{\mu\nu'}(x,x')\right).
\ee
It is convenient to work with the function
\be
\zeta=\frac{1}{2}\log \Delta\,.
\ee
It saisfies the equation
\be
\sigma^\mu\nabla_\mu\zeta=\frac{1}{2}\left(n-\sigma^\mu{}_\mu\right).
\label{zeta0}
\ee

We also introduce the operator 
of parallel transport ${\cal P}(x,x')$
of sections along the geodesic from the point $x'$
to the point $x$. 
It satisfies the equation
\be
\sigma^{\mu}\nabla_\mu{\cal P}=0.
\ee
It has the obvious properties
\be
{\cal P}^*(x,x')={\cal P}^{-1}(x,x')={\cal P}(x',x)\,.
\ee

We need to compute the mixed derivative of the heat kernel.
Since we want to study the asymptotics as $t\to 0$
we present it in the following form
\cite{dewitt65,avramidi00,avramidi91b}.
We fix a point $x'$
and consider a geodesic ball $B_r(x')$ centered at $x'$
of radius $r$ less than the injectivity radius $r_{\rm inj}(M)$
of the manifold, $r<r_{\rm inj}(M)$.
Then in this ball the heat kernel can be presented in the form
\be
U(t;x,x')=(4\pi t)^{-n/2}\exp\left\{-\frac{\sigma(x,x')}{2t}\right\}
\Psi(t;x,x'),
\ee
where
\be
\Psi(t;x,x')=\exp\left\{\zeta(x,x')\right\} {\cal P}(x,x')\Omega(t;x,x').
\ee
Here $\Omega$ is the so-called transfer function
that has the following asymptotic expansion
as $t\to 0$
\be
\Omega(t;x,x')\sim\sum_{k=0}^\infty \frac{(-t)^k}{k!}a_k(x,x')\,,
\ee
where $a_k$ are the so-called off-diagonal heat kernel
coefficients. In other words, the function $\Psi$
has the asymptotic expansion
\be
\Psi(t;x,x')\sim\sum_{k=0}^\infty t^k\psi_k(x,x'),
\label{56xx}
\ee
where
\be
\psi_k
=\frac{(-1)^k}{k!}e^\zeta{\cal P}a_k
\,.
\ee

By using the asymptotics of the heat kernel it is easy to 
obtain the asymptotics of the heat trace invariant
as $t\to 0$
\be
\Theta(t)\sim
(4\pi t)^{-n/2}\sum_{k=0}^\infty \frac{(-t)^k}{k!}A_k,
\ee
where
\be
A_k=\int\limits_M dv\, \tr[a_k],
\ee
where $[a_k]$ are the diagonal values of the heat kernel
coefficients. The first three coefficients are
(see \cite{gilkey95,avramidi91b,avramidi00})
\bea
A_0&=&N\vol(M)\,,
\\
A_1&=&\int\limits_Mdv\,\left(\tr Q-\frac{N}{6}R\right)\,,
\label{b1xx}
\\
A_2 &=&\int\limits_M dv\,
\Biggl\{\tr \left(Q^2-\frac{1}{3}QR
+\frac{1}{6}{\cal R}_{\mu\nu}{\cal R}^{\mu\nu}\right)
\nonumber\\
&&
+N\left(\frac{1}{36}R^2
-\frac{1}{90} R_{\mu\nu}R^{\mu\nu}
+\frac{1}{90}R_{\mu\nu\alpha\beta}R^{\mu\nu\alpha\beta}\right)
\Biggr\}
\,.
\label{b2xx}
\eea

It is pretty easy to compute the asymptotic expansion of the 
heat content $\Pi(t)$ for scalar operators
on manifolds without boundary.
First, note that
\bea
\int\limits_M dv' U(t;x,x') &=&
\left(\exp(-tL)\cdot 1\right)(x).
\eea
Therefore, the expansion of the heat content as $t\to 0$
has the form
\be
\Pi(t) =
\sum_{k=1}^\infty \frac{(-t)^k}{k!}
\Pi_k,
\ee
where
\bea
\Pi_0&=&\vol M,
\\
\Pi_1 &=&
\int\limits_M dv\,Q\,,
\\
\Pi_k &=&
\int\limits_M dv\,L^k\cdot 1
=\int\limits_M dv\,
Q\left(-\Delta+Q\right)^{k-2}Q\,, \qquad k\ge 2.
\eea
Notice that for pure Laplacian, when $Q=0$, 
the heat content is constant and is equal to its value at $t=0$,
$\Pi(t)=\vol(M)$.

\section{Mixed Derivative of the Heat Kernel}
\setcounter{equation}{0}

Now, we compute
\bea
\nabla_\mu\nabla_{\nu'}U &=& (4\pi t)^{-n/2}
\exp\left(-\frac{\sigma}{2t}\right)
\Biggl\{
\frac{1}{4t^2}\sigma_{\mu}\sigma_{\nu'}\Psi
\nonumber\\
&&
-\frac{1}{2t}\left(\sigma_{\mu\nu'}\Psi
+\sigma_\mu\Psi_{;\nu'}
+\sigma_{\nu'}\Psi_{;\mu}\right)
+\Psi_{;\mu\nu'}
\Biggr\}\,.
\eea
The derivatives of the function $\Psi$ read
\bea
\Psi_{;\mu} &=& e^\zeta
\left\{\left(\zeta_{;\mu} {\cal P}
+{\cal P}_{;\mu}\right)\Omega
+{\cal P}\Omega_{;\mu}\right\},
\\
\Psi_{;\mu'} &=& e^\zeta\left\{
\left(\zeta_{;\mu'}{\cal P}
+{\cal P}_{;\mu'}\right)\Omega
+{\cal P}\Omega_{;\mu'}\right\},
\\
\Psi_{;\mu\nu'} &=& e^\zeta
\Bigl\{
\left(\zeta_{;\mu\nu'}{\cal P}
+\zeta_{;\mu}\zeta_{;\nu'}{\cal P}
+\zeta_{;\mu}{\cal P}_{;\nu'}
+\zeta_{;\nu'}{\cal P}_{;\mu}
+{\cal P}_{;\mu\nu'}\right)\Omega
\nonumber\\
&&
+\zeta_{;\mu}{\cal P}\Omega_{;\nu'}
+\zeta_{;\nu'}{\cal P}\Omega_{;\mu}
+{\cal P}_{;\mu}\Omega_{;\nu'}
+{\cal P}_{;\nu'}\Omega_{;\mu}
+{\cal P}\Omega_{;\mu\nu'}\Bigr\}.
\eea

Therefore, we can write it in the form
\be
P_{\mu\nu'}=\tr U^*\nabla_\mu\nabla_{\nu'}U  
= \frac{1}{2t}
(4\pi t)^{-n}\exp\left(-\frac{\sigma}{t}\right)e^{2\zeta}
\Lambda
(-\sigma_{\mu\alpha'})Y^{\alpha'}{}_{\nu'},
\ee
where
\bea
\Lambda &=& \tr\Omega^*\Omega,
\\
Y^{\alpha'}{}_{\nu'} &=&
Z^{\alpha'}{}_{\nu'}
+W^{\alpha'}{}_{\nu'}
+V^{\alpha'}{}_{\nu'}
+\frac{1}{2t}S^{\alpha'}{}_{\nu'}\,,
\eea
with
\bea
Z^{\alpha'}{}_{\nu'}&=&\delta^{\alpha'}{}_{\nu'}
-2tF^{\alpha'}{}_{\nu'},
\label{65xx}
\\
V^{\alpha'}{}_{\nu'}&=&\sigma^{\alpha'} E_{\nu'},
\\
W^{\alpha'}{}_{\nu'}&=&\tilde E^{\alpha'}\sigma_{\nu'},
\\
S^{\alpha'}{}_{\nu'}&=&-\sigma^{\alpha'}\sigma_{\nu'}.
\eea
and
\bea
E_{\nu'}&=&\Lambda^{-1}{\cal E}_{\nu'},
\label{69xx}
\\
\tilde E^{\alpha'}&=&\Lambda^{-1}\gamma^{\alpha'\mu}
\tilde{\cal E}_\mu\,,
\\
F^{\alpha'}{}_{\nu'}&=&
\Lambda^{-1}
\gamma^{\alpha'\mu}{\cal F}_{\mu\nu'}.
\label{611xx}
\eea
The needed tensors have the form
\bea
{\cal E}_\mu &=&
\tr\Omega^*\left\{\left(\zeta_{;\mu} 
+{\cal P}^{-1}{\cal P}_{;\mu}\right)\Omega
+\Omega_{;\mu}\right\},
\\
\tilde{\cal E}_{\mu'} &=& \tr
\Omega^*\left\{\left(\zeta_{;\mu'}
+{\cal P}^{-1}{\cal P}_{;\mu'}\right)\Omega
+\Omega_{;\mu'}\right\},
\\
{\cal F}_{\mu\nu'} &=& 
\tr\Omega^*\Bigl\{
\left(\zeta_{;\mu\nu'}
+\zeta_{;\mu}\zeta_{;\nu'}
+\zeta_{;\mu}{\cal P}^{-1}{\cal P}_{;\nu'}
+\zeta_{;\nu'}{\cal P}^{-1}{\cal P}_{;\mu}
+{\cal P}^{-1}{\cal P}_{;\mu\nu'}\right)\Omega
\nonumber\\
&&
+\zeta_\mu\Omega_{;\nu'}
+\zeta_{;\nu'}\Omega_{;\mu}
+{\cal P}^{-1}{\cal P}_{;\mu}\Omega_{;\nu'}
+{\cal P}^{-1}{\cal P}_{;\nu'}\Omega_{;\mu}
+\Omega_{;\mu\nu'}\Bigr\}.
\eea

Now, we need to compute the determinant of the matrix $P=P_{\mu\nu'}$.
We obtain
\be
\det P=g^{1/2}(x)g^{1/2}(x')(2t)^{-n}(4\pi t)^{-n^2}
\exp\left(-n\frac{\sigma}{t}\right)\Delta H,
\ee
where
\bea
H &=&e^{2n\zeta}\Lambda^n\det Y^{\alpha'}{}_{\nu'}.
\eea

We will use the following formula for the determinant of the
sum of two matrices.
By using the following formula for the determinant of the matrix
$I+C$,
\bea
\det(I+C) &=&
\sum_{k=0}^{n}
C^{\alpha_1}{}_{[\alpha_1}\cdots C^{\alpha_{k}}{}_{\alpha_{k}]}
\nonumber\\
&=&
1+\tr C+\frac{1}{2}(\tr C)^2-\frac{1}{2}\tr C^2+\cdots+
C^{\alpha_1}{}_{[\alpha_1}\cdots C^{\alpha_{n}}{}_{\alpha_{n}]}
\eea
we obtain the determinant of the sum of two matrices
$A+B$ (with $B$ being an invertible matrix)
\bea
\det(A+B) &=&
\det B
\sum_{k=0}^{n}
A^{\beta_1}{}_{[\alpha_1}\cdots A^{\beta_{k}}{}_{\alpha_{k}]}
B^{-1}{}^{\alpha_1}{}_{\beta_1}\cdots
B^{-1}{}^{\alpha_k}{}_{\beta_k}.
\label{615xx}
\eea

Let $X=(X^{\nu'}{}_{\mu'})$ be the inverse of the matrix $Z=(Z^{\mu'}{}_{\nu'})$
and
\be
J=\det Z\,.
\label{616xx}
\ee
Next, we note that
\be
S^{[\mu_1}{}_{[\nu'_1}S^{\mu_2]}{}_{\nu'_2]}
=V^{[\mu_1}{}_{[\nu'_1}V^{\mu_2]}{}_{\nu'_2]}
=W^{[\mu_1}{}_{[\nu'_1}W^{\mu_2]}{}_{\nu'_2]}
=S^{[\mu_1}{}_{[\nu'_1}V^{\mu_2]}{}_{\nu'_2]}
=S^{[\mu_1}{}_{[\nu'_1}W^{\mu_2]}{}_{\nu'_2]}
=0.
\ee
Then
by using the above formula (\ref{615xx})
for the determinant we obtain
\be
H = Je^{2n\zeta}\Lambda^n\left(1+V^{\mu'}{}_{\nu'}X^{\nu'}{}_{\mu'}
+W^{\mu'}{}_{\nu'}X^{\nu'}{}_{\mu'}
+\frac{1}{2t}S^{\mu'}{}_{\nu'}X^{\nu'}{}_{\mu'}
+2V^{[\mu'}{}_{\alpha'}W^{\nu']}{}_{\beta'}X^{\alpha'}{}_{\mu'}
X^{\beta'}{}_{\nu'}\right)
\ee
This can be expressed in terms of a few invariants; let
\bea
\chi_1 &=&\sigma_{\beta'}X^{\beta'}{}_{\nu'}\sigma^{\nu'},
\\
\chi_2&=&E_{\nu'}X^{\nu'}{}_{\mu'}\sigma^{\mu'},
\\
\chi_3&=& \sigma_{\nu'}X^{\nu'}{}_{\mu'}\tilde E^{\mu'},
\\
\chi_4&=&E_{\alpha'}X^{\alpha'}{}_{\mu'}\tilde E^{\mu'}.
\eea
Then
\bea
H&=& Je^{2n\zeta}\Lambda^n\left(
-\frac{1}{2t}\chi_1+1+\chi_2+\chi_3
+\chi_2\chi_3-\chi_1\chi_4
\right).
\eea

\section{Heat Determinant Asymptotics}
\setcounter{equation}{0}

We are studying the asymptotics as $t\to 0$ of the functional
$K(t)$. As we mentioned above we fix a point $x'$ in the manifold $M$
and consider a geodesic ball $B_r(x')$ centered at $x'$ of radius 
$r<r_{\rm inj}(M)$
smaller than the injectivity radius $r_{\rm inj}(M)$
of the manifold. We decompose the integral defining the invariant $K(t)$
in two parts
\be
K(t)=K_{\rm diag}(t)+K_{\rm off-diag}(t),
\ee
where
\bea
K_{\rm diag}(t)&=&
\int\limits_{M}dx'\,\int\limits_{B_r(x')}dx\,
\det P_{\mu\nu'}(t;x,x'),
\\
K_{\rm off-diag}(t)&=&\int\limits_M dx' \int\limits_{M-B_r(x')}dx\;
\det P_{\mu\nu'}(t;x,x')\,.
\eea

By using the standard elliptic estimates of the heat kernel
\cite{grigoryan09}
one can show that for any $x\in M-B_r(x,x')$ and $0<t<1$
there is an estimate
\be
\left|U(t;x,x')\right|\le C_1 t^{-n/2}\exp\left(-\frac{r^2}{4t}\right),
\ee
and similarly
\be
\left|P_{\mu\nu'}(t;x,x')\right|\le 
C_2 t^{-n-2}\exp\left(-\frac{r^2}{2t}\right),
\ee
where $C_1$ and $C_2$ are constants; therefore,
\be
\left|\det P_{\mu\nu'}(t;x,x')\right|\le 
C_2 t^{-n(n+2)}\exp\left(-\frac{nr^2}{2t}\right),
\ee
and
\be
\left|K_{\rm off-diag}(t)\right|\le 
C_3 t^{-n(n+2)}\exp\left(-\frac{nr^2}{2t}\right),
\ee
where $C_3$ is another constant.

Thus, we see that the off-diagonal part $K_{\rm off-diag}(t)$
is exponentially small as $t\to 0$ and does not contribute
to the asymptotic expansion of the invariant $K(t)$ as $t\to 0$,
that is, as $t\to 0$
\be
K(t)\sim K_{\rm diag}(t)\,,
\ee
and, hence,
\be
K(t) 
\sim 2^{-n}(4\pi)^{-n^2}
t^{-n(n+1)}\int\limits_{M}dv'\,\int\limits_{B_r(x')}dv\,
\exp\left(-n\frac{\sigma(x,x')}{t}\right)
\Delta(x,x')H(t;x,x')\,.
\ee

To compute this integral as $t\to 0$
we make the change of variables
$x^\mu\mapsto \xi^{\nu'}$ described below.
It is convenient to introduce new coordinates
\be
\xi^{\nu'}=\frac{\sigma^{\nu'}}{\sqrt{t}}\,,
\ee
Note that
\be
|\xi|^2=\xi^{\nu'}\xi_{\nu'}
=\frac{2\sigma}{t}.
\ee
The volume element changes as follows
\be
dv(x)=g^{1/2}(x)\;dx = t^{n/2}\Delta^{-1}(x,x') g^{1/2}(x') d\xi\,.
\ee
The integration over $\xi$ goes over the Euclidean ball $B_{r/\sqrt{t}}(0)$
of radius $r/\sqrt{t}$, that is, $|\xi|\le r/\sqrt{t}$.
As $t\to 0$ the radius of the ball goes to infinity, therefore,
the asymptotics $t\to 0$ are determined by the intergal over the whole
Euclidean space $\RR^n$. Alternatively, we could split the intergral over
the ball $B_{r/\sqrt{t}}(0)$ as the integral over the whole $\RR^n$
minus the integral over $\RR^n-B_{r/\sqrt{t}}(0)$ and then
show that the second integral is exponentially small
as $t\to 0$.

Thus, we get 
\bea
K(t)
&\sim&  2^{-n}(4\pi)^{-n^2}
t^{-n\left(n+\frac{1}{2}\right)}
\nonumber\\
&&
\times\int\limits_{M} dv'
\int\limits_{\RR^n} d\xi\;g^{1/2}(x')\;
\exp\left(-\frac{n}{2}|\xi|^2\right)
H(t;x,x')\,.
\eea
This can be rewritten in the form
\be
K(t) \sim (4\pi)^{-n^2}\left(\frac{\pi}{2n}\right)^{n/2}
t^{-n\left(n+\frac{1}{2}\right)}
\int\limits_{M}dv'\,
\left<H\right>\,,
\ee
where the brackets $\left<\dots\right>$ denote
the Gaussian average over the variables $\xi$
defined by
\bea
\left<f(\xi)\right>
&=&
\left(\frac{n}{2\pi}\right)^{n/2}
\int\limits_{\RR^n}d\xi\, g^{1/2}(x')\;
\exp\left(-\frac{n}{2}|\xi|^2\right)f(\xi)
\,.
\eea

By using the asymptotic expansion
(\ref{56xx}) one can show that
all matrices $Z, V, W, S$ introduced above have
asymptotic expansions in non-negative integer powers of $t$;
therefore, there exists an expansion
\be
H(t;x,x')\sim \sum_{k=-1}^\infty t^{k} h_{k}(x,x').
\ee
Therefore,
\be
K(t) \sim (4\pi)^{-n^2}\left(\frac{\pi}{2n}\right)^{n/2}
t^{-n\left(n+\frac{1}{2}\right)}
\sum_{k=-1}^\infty
H_k(t)\,,
\ee
where
\be
H_k(t) =
\int\limits_{M}dv'\,\left<h_k\right>\,.
\ee

To compute the Gaussian average we expand the coefficients $h_k$
in covariant Taylor series at the point $x'$
\bea
h_k(x,x')
&=&\sum_{m=0}^\infty
\frac{(-1)^m}{m!}
\sigma^{\mu'_1}\cdots\sigma^{\mu'_m}h_{k,\mu'_1\dots\mu'_m}(x')
\nonumber\\
&=&\sum_{m=0}^\infty t^{m/2}\frac{(-1)^m}{m!}
\xi^{\mu'_1}\cdots\xi^{\mu'_m}h_{k,\mu'_1\dots\mu'_m}(x')
\,,
\eea
where
\be
h_{k,\mu'_1\dots\mu'_m}(x')
= \left[\nabla_{(\mu_1}\cdots\nabla_{\mu_m)}h_k(x,x')\right]_{x=x'}\,.
\ee
The Gaussian averages of the monomials are well known
\bea
\left<\xi^{\mu_1}\cdots \xi^{\mu_{2k+1}}\right>&=&0\,,
\\[10pt]
\left<\xi^{\mu_1}\cdots \xi^{\mu_{2k}}\right>
&=&\frac{(2k)!}{ (2n)^{k}k!}\,
g^{(\mu_1\,\mu_2}\cdots g^{\mu_{2k-1}\,\mu_{2k})}\,.
\eea
In particular,
\bea
\left<\xi^{\mu}\xi^{\nu}\right>
&=&\frac{1}{n}\,g^{\mu\nu}\,,
\\
\left<\xi^{\mu}\xi^\nu\xi^\alpha\xi^\beta\right>
&=&\frac{1}{n^2}\,
\left(g^{\mu\nu}g^{\alpha\beta}
+g^{\mu\alpha}g^{\nu\beta}
+g^{\mu\beta}g^{\nu\alpha}\right)
\,.
\eea

This enables us to immediately compute the Gaussian average
\be
\left<h_k\right>
=\sum_{j=0}^\infty t^j h_{k,j}\,,
\ee
where
\be
h_{k,j}=\frac{1}{(2n)^{j}j!}\,
g^{\mu_1\,\mu_2}\cdots g^{\mu_{2j-1}\,\mu_{2j}}
h_{k,\mu'_1\dots\mu'_{2j}}\,.
\ee
So,
\be
H_k(t)=\sum_{j=0}^\infty t^j H_{k,j}\,,
\ee
where
\be
H_{k,j}=\int\limits_{M}dv\,h_{k,j}\,.
\ee

Therefore, the heat determinant has the asymptotics
as $t\to 0$
\be
K(t)\sim (4\pi)^{-n^2}
\left(\frac{\pi}{2n}\right)^{n/2}
t^{-n\left(n+\frac{1}{2}\right)}
\sum_{k=-1}^\infty
t^k B_k,
\ee
where
\be
B_k=\int\limits_{M}dv\,b_k
=\sum_{j=-1}^kH_{j,k-j},
\ee
with
\be
b_k=\sum_{j=-1}^kh_{j,k-j}\,.
\ee

The coefficients $B_k$ are the new heat invariants 
that are of central interest of this paper.
They are not spectral invariants since they depend on the
eigenfunctions as well. 
They are invariants built from the curvatures and their derivatives
(as well as the potential term $Q$) with numerical coefficients
that are universal since they
depend only on the dimension of the manifold and the dimension 
of the vector bundle.
Their calculation is reduced to the calculation of the coefficients
$H_{k,m}$, which, in turn, are determined by the Taylor
coefficients $h_{k,\mu'_1\dots\mu'_{2m}}$
of the coefficients $h_k$.

\section{Calculation of Low Order Coefficients}
\setcounter{equation}0

\subsection{Calculation of the Coefficients $h_-1$, $h_0$, $h_1$, $h_2$}

We will compute the first four coefficients 
\bea
b_{-1} &=& h_{-1,0},
\\
b_0 &=& h_{0,0}+h_{-1,1},
\\
b_1 &=& h_{1,0}+h_{0,1}+h_{-1,2},
\\
b_2 &=& h_{2,0}+h_{1,1}+h_{0,2}+h_{-1,3}.
\eea
We have
\bea
h_{-1,0} &=& [h_{-1}],
\\
h_{-1,1} &=& \frac{1}{2n}g^{\mu\nu}h_{-1, \mu\nu},
\\
h_{-1,2} &=& \frac{1}{8n^2}
g^{\mu\nu}g^{\alpha\beta}h_{-1, \mu\nu\alpha\beta},
\\
h_{-1,3} &=& \frac{1}{48n^3}
g^{\mu\nu}g^{\alpha\beta}g^{\rho\sigma}
h_{-1, \mu\nu\alpha\beta\rho\sigma},
\\
h_{0,0} &=& [h_0],
\\
h_{0,1} &=& \frac{1}{2n}g^{\mu\nu}h_{0, \mu\nu},
\\
h_{0,2} &=& \frac{1}{8n^2}
g^{\mu\nu}g^{\alpha\beta}h_{0,\mu\nu\alpha\beta},
\eea
\bea
h_{1,0} &=& [h_1],
\\
h_{1,1} &=& \frac{1}{2n}g^{\mu\nu}h_{1, \mu\nu},
\\
h_{2,0} &=& [h_2].
\eea 
Here the square brackets denote the coincidence limits, as usual.

Thus, we need to compute first the coefficients $h_{-1}, h_0, h_1, h_2$
\be
H\sim t^{-1}h_{-1}+h_0+t h_1+t^2h_2+\cdots
\ee
in terms of $\psi_k$
and then compute the coincidence limits of their derivatives
(or expand them in Taylor series).
Note that we need $h_{-1}$ up to sixth order in $\sigma^{\mu'}$,
the coefficient $h_0$ up to the forth order in $\sigma^{\mu'}$
the coefficient $h_1$ up to the second order and
the coefficient $h_2$ up to the zero order. 
What we actually do is introduce a small parameter $\varepsilon$
so that
\be
t\sim \varepsilon^2, \qquad 
\sigma^{\mu'}\sim \varepsilon.
\ee 
Since we want to compute the terms of the order $t^2$
we need to keep the terms of the order $\varepsilon^4$
and neglect the terms of higher order in $\varepsilon$.
All expansions below are valid in this approximation.
The dots below denote the neglected terms of order
$O(\varepsilon^5)$.

We have
\be
H=Je^{2n\zeta}\Lambda^n\Phi,
\ee
where
\be
\Phi=-\frac{1}{2t}\chi_1+1+\chi_2+\chi_3+\chi_2\chi_3
-\chi_1\chi_4\,.
\ee

The expansion of the function $\Lambda$ has the form
\be
\Lambda =N+t\Lambda_1+t^2\Lambda_2+\cdots,
\ee
so that
\be
\Lambda^{-1}=\frac{1}{N}-\frac{1}{N^2}t\Lambda_1
+t^2\left(\frac{1}{N^3}\Lambda_1^2-\frac{1}{N^2}\Lambda_2\right)
+\cdots
\ee
and
\be
\Lambda^n =N^n+tnN^{n-1}\Lambda_1
+t^2\left(nN^{n-1}\Lambda_2+\frac{n(n-1)}{2}N^{n-2}\Lambda_1^2\right)+\cdots,
\ee
where
\bea
\Lambda_1&=& -\tr(a_1+a^*_1)\,,
\\
\Lambda_2&=&\frac{1}{2}
\tr\left(a_2+a_2^*+2a_1a_1^*\right)\,.
\eea

To compute the expansion of needed functions we will need
to find the expansion of the tensors $E$, $\tilde E$ and $F$
(\ref{69xx})-(\ref{611xx}). We will compute them later.
For now, let us denote them by
\bea
E_{\nu'}&=& E_{0, \nu'}+tE_{1,\nu'}+t^2E_{2,\nu'}+\cdots,
\\
\tilde E^{\nu'}&=& \tilde E^{\nu'}_{0}+tE^{\nu'}_{1}+t^2E^{\nu'}_{2}+\cdots,
\\
F^{\alpha'}{}_{\nu'}&=& F_{0}^{\alpha'}{}_{\nu'}
+tF_{1}^{\alpha'}{}_{\nu'}+t^2F_{2}^{\alpha'}{}_{\nu'}+\cdots,
\eea
Then the expansion of the function $J$ 
(\ref{616xx}), the determinant of the matrix $Z$ (\ref{65xx}),
has the form
\be
J =1+tJ_1+t^2 J_2+\cdots,
\ee
where
\bea
J_1&=& -2F_0^{\mu'}{}_{\mu'}\,,
\\
J_2&=& 2F_0^{\mu'}{}_{\mu'}F_0^{\nu'}{}_{\nu'}
-2F_0^{\mu'}{}_{\nu'}F_0^{\nu'}{}_{\mu'}-2F_{1}^{\mu'}{}_{\mu'}
\,.
\eea
The expansion of the matrix $X$, the inverse of the matrix $Z$ (\ref{65xx}), reads
\be
X^{\alpha'}{}_{\nu'}=\delta^{\alpha'}{}_{\nu'}
+t X_1^{\alpha'}{}_{\nu'}+t^2 X_2^{\alpha'}{}_{\nu'}+\dots\,,
\ee
where
\bea
X_1^{\alpha'}{}_{\nu'}&=&2F_0^{\alpha'}{}_{\nu'},
\\
X_2^{\alpha'}{}_{\nu'}&=&4F_0^{\alpha'}{}_{\mu'}F_0^{\mu'}{}_{\nu'}
+2F_1^{\alpha'}{}_{\nu'}\,.
\eea

We need to compute the expansion of the function $\Phi$ up to
terms of order $O(\varepsilon^5)$ (recall that $t\sim \varepsilon^2$
and $\sigma^{\mu'}\sim \varepsilon$).
Note that
\be
\chi_1 \sim O(\varepsilon^2), \qquad
\chi_2, \chi_{3} \sim O(\varepsilon), \qquad
\chi_{4}\sim O(1).
\ee
Therefore, since $\chi_1$ is divided by $t$ we need $\chi_1$ 
up to terms of order $t^2$.
Also, we only need to keep terms up to order $t$ in 
$\chi_2$ and $\chi_3$.
Further, since the function $\chi_4$ comes only with the product
with the function $\chi_1$, we need the function $\chi_4$ also
only up to linear terms in $t$, that is
\bea
\chi_1 &=& \chi_{1,0}+t\chi_{1,1}+t^2\chi_{1,2}\cdots,
\\
\chi_2 &=& \chi_{2,0}+t\chi_{2,1}+\cdots,
\\
\chi_3 &=& \chi_{3,0}+t\chi_{3,1}+\cdots,
\\
\chi_4 &=& \chi_{4,0}+t\chi_{4,1}+\cdots.
\eea
By using the expansion of the matrix $X$ and the quantities
$E$, $\tilde E$ and $F$ we obtain
\bea
\chi_{1,0} &=&2\sigma,
\\
\chi_{1,1} &=&2\sigma_{\nu'}F_0^{\nu'}{}_{\mu'}\sigma^{\mu'},
\\
\chi_{1,2} &=&2\sigma_{\nu'}F_1^{\nu'}{}_{\mu'}\sigma^{\mu'}
+4\sigma_{\nu'}F_0^{\nu'}{}_{\alpha'}F_0^{\alpha'}{}_{\mu'}\sigma^{\mu'},
\\
\chi_{2,0}&=&E_{0,\mu'}\sigma^{\mu'},
\\
\chi_{2,1}&=&2E_{1,\mu'}\sigma^{\mu'}
+2E_{0,\nu'}F_0^{\nu'}{}_{\mu'}\sigma^{\mu'},
\\
\chi_{3,0}&=& \sigma_{\mu'}\tilde E_0^{\mu'},
\\
\chi_{3,1}&=&2\sigma_{\nu'}\tilde E_1^{\mu'}
+2\sigma_{\nu'}F_0^{\nu'}{}_{\mu'}\tilde E_0^{\mu'},
\\
\chi_{4,0}&=&E_{0,\mu'}\tilde E_0^{\mu'}.
\\
\chi_{4,1}&=&
E_{0,\mu'}\tilde E_1^{\mu'}
+E_{1,\mu'}\tilde E_0^{\mu'}
+2E_{0,\nu'}F_0^{\nu'}{}_{\mu'}\tilde E_0^{\mu'}.
\eea

By the same reason 
we only need the expansion of the 
function $\Phi$ up to linear terms in $t$
(we neglect the terms of order higher than $O(\varepsilon^4)$;
the quadratic terms in $t$ will be of order $O(\varepsilon^5)$).
We get
\be
\Phi =-\frac{\sigma}{t}+
\Phi_0+t\Phi_1+\cdots,
\ee
where
\bea
\Phi_0&=&
1-\frac{1}{2}\chi_{1,1}+\chi_{2,0}
+\chi_{3,0}+\chi_{2,0}\chi_{3,0}
-\chi_{1,0}\chi_{4,0}\,,
\\
\Phi_1&=&
-\frac{1}{2}\chi_{1,2}+\chi_{2,1}+\chi_{3,1}
+\chi_{2,0}\chi_{3,1}+\chi_{2,1}\chi_{3,0}
-\chi_{1,1}\chi_{4,0}-\chi_{1,0}\chi_{4,1}
\,.
\eea

This gives the coefficients $h_k$ of the expansion of the function $H$
(in the needed order in $\varepsilon$)
\bea
h_{-1} &=& -N^n \sigma e^{2n\zeta}\,,
\\
h_0 &=& e^{2n\zeta}\left\{N^n\Phi_0-N^n\sigma J_1-nN^{n-1}\sigma\Lambda_1\right\},
\\
h_1 &=&e^{2n\zeta}\Biggl\{N^n J_1\Phi_0+nN^{n-1}\Lambda_1\Phi_0+N^n\Phi_1
\nonumber\\
&&
-\sigma\left(N^nJ_2
+nN^{n-1}\Lambda_2
+\frac{n(n-1)}{2}N^{n-2}\Lambda_1^2
+nN^{n-1}J_1\Lambda_1
\right)\Biggr\},
\\
h_2 &=&e^{2n\zeta}\Biggl\{
N^nJ_1\Phi_1+nN^{n-1}\Lambda_1\Phi_1
\nonumber\\
&&
+\Phi_0\left(N^nJ_2
+nN^{n-1}\Lambda_2
+\frac{n(n-1)}{2}N^{n-2}\Lambda_1^2
+nN^{n-1}J_1\Lambda_1
\right)\Biggr\}.
\eea

\subsection{Calculation of $b_{-1}$}

By using the explicit form of the coefficient $h_{-1}$ we
see that its coincidence limit vanishes, 
\be
h_{-1,0} =[h_{-1}]= 0\,.
\ee
Thus, the coefficient $b_{-1}$ vanishes as well
\be
b_{-1}=0\,.
\ee

\subsection{Calculation of $b_0$}

Next, by using the coefficient $h_{-1}$ 
and the coincidence limits of the 
functions $\sigma$ and $\zeta$ (see Appendix)
we compute
\be
h_{-1,\mu\nu}=-N^n g_{\mu\nu}\,.
\ee
Therefore,
\be
h_{-1,1} = -\frac{1}{2}N^n\,.
\ee

Next, we compute
\be
h_{0,0}=N^n[\Phi_0]\,,
\ee
It is easy to see that
\be
[\Phi_0]=1,
\ee
and, therefore,
\be
h_{0,0}=N^n\,.
\ee
This gives the coefficient $b_0$
\be
b_0=\frac{1}{2}N^n.
\ee

\subsection{Calculation of $b_1$}

First, by using the derivatives of the function $\zeta$
we obtain
\be
h_{-1,\mu\nu\alpha\beta}
=-\frac{n}{3}N^ng_{(\mu\nu}R_{\alpha\beta)}\,.
\ee
By contracting all indices we get
\bea
h_{-1,2} &=& -N^n\frac{n+2}{72n}R\,.
\eea

Then we compute
\be
h_{1,0}=N^n [J_1]+nN^{n-1}[\Lambda_1]+N^n[\Phi_1].
\ee
We have
\be
[\Lambda_1]=-\tr [a_1+a_1^*]
=-2\tr Q +\frac{1}{3}NR\,
\ee
and
\be
[\Phi_1]=0\,.
\ee
To compute 
\be
[J_1]=-2[F_{0}^{\mu'}{}_{\mu'}],
\ee
we need to compute the the coincidence limit $[F_{0}^{\mu'}{}_{\mu'}]$.
By using the coincidence limits of the derivatives of the functions
$\zeta$ and ${\cal P}$ we obtain
\be
[F_0^{\alpha'}{}_{\nu'}]
=\frac{1}{6}R^\alpha{}_\nu.
\ee
Therefore,
\be
[J_1]=-\frac{1}{3}R
\ee
and
\be
h_{1,0}=N^n \frac{(n-1)}{3}R-2nN^{n-1}\tr Q.
\ee

Now we need to compute $h_{0,1}$. First, we have
\be
h_{0,\mu\nu}=2nN^n[\zeta_{;\mu\nu}]
+N^n [\Phi_{0;(\mu\nu)}]-N^n [J_1]g_{\mu\nu}
-nN^{n-1}[\Lambda_1] g_{\mu\nu}\,.
\ee
Thus we need to compute $[\Phi_{0,\mu\nu}]$.
By using the definition of $\Phi_0$ and the
functions $\chi_k$ we get
\be
[\Phi_{0;(\mu\nu)}] =
\left[-\frac{1}{2}\chi_{1,1;(\mu\nu)}
+\chi_{2,0;(\mu\nu)}
+\chi_{3,0;(\mu\nu)}
+2\chi_{2,0;(\mu}\chi_{3,0;\nu)}
-\chi_{1,0;(\mu\nu)}\chi_{4,0}\right]\,.
\ee
This gives
\bea
[\chi_{1,0;(\mu\nu)}] &=&2g_{\mu\nu},
\\
{}[\chi_{1,1;(\mu\nu)}] &=&[4F_{0,(\mu'\nu')}]
=\frac{2}{3}R_{\mu\nu},
\\
{}[\chi_{2,0;\mu}]&=&-[E_{0,\mu'}],
\\
{}[\chi_{3,0;\mu}]&=& -[\tilde E_{0,\mu'}],
\\
{}[\chi_{2,0;(\mu\nu)}]&=&-2[E_{0,(\mu';\nu)}],
\\
{}[\chi_{3,0;(\mu\nu)}]&=& -2[\tilde E_{0,(\mu';\nu)}],
\\
{}[\chi_{4,0}]&=&[E_{0,\mu'}\tilde E_0^{\mu'}].
\eea

Now, we need to compute the coincidence limits of the
functions $E_0$ and $\tilde E_0$
and their derivatives.
It is easy to see that
\be
[E_{0,\mu'}]=[\tilde E_{0,\mu'}]=0\,.
\ee
So,
\be
[\chi_{2,0,\mu}]=[\chi_{3,0,\mu}]=[\chi_{4,0}]=0.
\ee
Therefore,
\be
[\Phi_{0;(\mu\nu)}] =
\left[-\frac{1}{2}\chi_{1,1;(\mu\nu)}
+\chi_{2,0;(\mu\nu)}
+\chi_{3,0;(\mu\nu)}\right]\,.
\ee
Next, by using the coincidence limits of the functions $\zeta$
and ${\cal P}$ we get
\bea
[\tilde E_{0,(\mu';\nu)}] &=& -[\zeta_{;\mu\nu}]
=-\frac{1}{6}R_{\mu\nu}\,,
\\
{}[E_{0,(\mu';\nu)}] &=& [\zeta_{;(\mu'\nu)}]
=-\frac{1}{6}R_{\mu\nu}\,.
\eea
This gives
\be
[\Phi_{0;(\mu\nu)}] = \frac{1}{3}R_{\mu\nu}\,.
\ee
Therefore,
\be
h_{0,\mu\nu}=N^n\frac{n+1}{3}R_{\mu\nu}
+N^{n}\frac{1-n}{3}R g_{\mu\nu}
+2nN^{n-1}g_{\mu\nu}\tr Q 
\,.
\ee
Thus, by contracting the indices we get
\be
h_{0,1}=\frac{1}{2n}g^{\mu\nu}h_{0,\mu\nu}
=N^n\frac{1+2n-n^2}{6n}R
+nN^{n-1}\tr Q 
\,.
\ee
Finally, by collecting all terms we get
\be
b_1=N^n\frac{12n^2-n+10}{72n}R
-nN^{n-1}\tr Q.
\ee

\subsection{Calculation of $b_2$}

To compute the coefficient $b_2$ we need to compute
$h_{-1,3}$, $h_{0,2}$, $h_{1,1}$ and $h_{2,0}$.

First, we compute
\bea
h_{-1,\mu_1\mu_2\mu_3\mu_4\mu_5\mu_6}
=-30nN^n\left(\zeta_{;(\mu_1\mu_2\mu_3\mu_4}g_{\mu_5\mu_6)}
+6n\zeta_{;(\mu_1\mu_2}\zeta_{;\mu_3\mu_4}g_{\mu_5\mu_6)}
\right)\,.
\eea

Next, we need to contract all indices.
One can show that for any two symmetric tensors $A$ and $B$
\be
g^{\mu_1\mu_2}g^{\mu_3\mu_4}g^{\mu_5\mu_6}A_{(\mu_1\mu_2\mu_3\mu_4}B_{\mu_5\mu_6)}
=\frac{1}{15}\left(
3A^{\mu\nu}{}_{\mu\nu}B^\alpha{}_\alpha
+12A^{\alpha}{}_{\alpha\mu\nu}B^{\mu\nu}\right).
\ee
Similarly, we can show that for any symmetric tensor $C$ we have
\be
g^{\mu_1\mu_2}C_{(\mu_1\mu_2}C_{\mu_3\mu_4)}
=\frac{1}{3}\left(C^{\mu}{}_{\mu}C_{\mu_3\mu_4}
+2C^{\alpha}{}_{(\mu_3}C_{\mu_4)\alpha}\right).
\ee
By using these  two equations we show that for 
any two symmetric tensors
$C$ and $B$ 
\bea
&&g^{\mu_1\mu_2}g^{\mu_3\mu_4}g^{\mu_5\mu_6}
C_{(\mu_1\mu_2}C_{\mu_3\mu_4}B_{\mu_5\mu_6)}
\\
&=&\frac{1}{15}\Biggl\{C^{\mu}{}_{\mu}C^\nu{}_\nu B^\alpha{}_\alpha
+2C^{\mu\nu}C_{\mu\nu} B^\alpha{}_\alpha
+4C^{\alpha}{}_{\alpha}C_{\mu\nu} B^{\mu\nu}
+8C^\alpha{}_{\mu}C_{\alpha\nu} B^{\mu\nu}
\Biggr\}.
\nonumber
\eea
This enables us to compute
\be
g^{\mu_1\mu_2}g^{\mu_3\mu_4}g^{\mu_5\mu_6}
\zeta_{;(\mu_1\mu_2\mu_3\mu_4}g_{\mu_5\mu_6)}
=\frac{n+4}{5}\zeta_{(4)},
\ee
where
\be
\zeta_{(4)}=g^{\mu_1\mu_2}g^{\mu_3\mu_4}\zeta_{;(\mu_1\mu_2\mu_3\mu_4)},
\ee
and
\bea
g^{\mu_1\mu_2}g^{\mu_3\mu_4}g^{\mu_5\mu_6}
\zeta_{;(\mu_1\mu_2}\zeta_{;\mu_3\mu_4}g_{\mu_5\mu_6)}
&=&\frac{n+4}{15}\left(\zeta^{;\mu}{}_{\mu}\zeta^{;\nu}{}_\nu
+2\zeta^{;\mu\nu}\zeta_{;\mu\nu}
\right).
\eea
Thus, we get
\bea
h_{-1,3}=-N^n\frac{(n+4)}{8n^2}
\left[\zeta_{(4)}
+2n\zeta^{;\mu}{}_{\mu}\zeta^{;\nu}{}_\nu
+4n\zeta^{;\mu\nu}\zeta_{;\mu\nu}
\right].
\eea
The tensors $[\zeta_{;\mu\nu}]$ and $[\zeta_{;(\mu\nu\alpha\beta)}]$
are listed in the Appendix.
By using these tensors we obtain
\bea
h_{-1,3}=-N^n\frac{(n+4)}{8n^2}
\left(
\frac{1}{5}\nabla_\mu\nabla^\mu R
+\frac{n}{18}R^2
+\frac{5n+1}{45}R_{\mu\nu}R^{\mu\nu}
+\frac{1}{30}R_{\mu\nu\alpha\beta}R^{\mu\nu\alpha\beta}
\right).
\eea

Next, we compute $h_{2,0}$. We already have 
\bea
[\Phi_1]&=&0,
\\
{}[J_1]&=&-\frac{1}{3}R,
\\
{}[\Lambda_1]&=& \frac{1}{3}NR-2\tr Q,
\\
{}[\Lambda_2]&=& \tr [a_2]+\tr [a_1]^2.
\eea
We compute $[J_2]$. 
By using $F_0$ we get
\be
[J_2]=\frac{1}{18}R^2-\frac{1}{18}R_{\mu\nu}R^{\mu\nu}
-2[F^{\mu'}_{1,\mu'}]\,.
\ee

Next, we compute the matrix $F_1$,
\be
[F^{\mu'}_{1,\nu'}]=\frac{1}{N}\tr [\nabla_{\nu'}\nabla^\mu a_{1}]
+\frac{2}{N}\tr[{\cal P}^{;\mu}{}_{\nu'}]Q.
\ee
The contraction of this matrix reads
\be
[F^{\mu'}_{1,\mu'}]=\frac{1}{N}\tr [\nabla_{\mu'}\nabla^\mu a_{1}].
\ee
We use the equation
\be
[\nabla^{\mu'}\nabla_\mu a_1]=\nabla^\mu[\nabla_\mu a_1]-[\nabla^\mu\nabla_\mu a_1]\,.
\ee
By using the equations in the appendix we obtain
\bea
[\nabla^{\mu'}\nabla_\mu a_1] &=&
\frac{1}{6}\nabla^\mu\nabla_{\mu}Q
+\frac{1}{6}{\cal R}_{\alpha \mu}{\cal R}^{\alpha\mu}
-\frac{1}{60}\nabla^\mu\nabla_\mu R
\\[5pt]
&&
-\frac{1}{90}R_{\mu\nu}R^{\mu\nu}
+\frac{1}{90}R_{\mu\nu\alpha\beta}R^{\mu\nu\alpha\beta}
\,.
\eea
This gives
\bea
[J_2] &=& 
\frac{1}{30}\nabla^\mu\nabla_\mu R
+\frac{1}{18}R^2
-\frac{1}{30}R_{\mu\nu}R^{\mu\nu}
-\frac{1}{45}R_{\mu\nu\alpha\beta}R^{\mu\nu\alpha\beta}
\nonumber
\\
&&
+\frac{1}{N}
\tr\left(
-\frac{1}{3}\nabla^\mu\nabla_{\mu}Q
-\frac{1}{3}{\cal R}_{\alpha \mu}{\cal R}^{\alpha\mu}
\right)
\,.
\eea

By collecting all terms we obtain
\bea
h_{2,0}&=&
-\frac{n+1}{3}N^{n-1}\tr\nabla^\mu\nabla_{\mu}Q
+\frac{2n+1}{30}N^n\nabla^\mu\nabla_\mu R
+\frac{n-2}{6}N^{n-1}\tr {\cal R}_{\mu\nu}{\cal R}^{\mu\nu}
\nonumber\\
&&
+2nN^{n-1}\tr Q^2
+2n(n-1)N^{n-2}(\tr Q)^2
-\frac{2n(n-1)}{3}N^{n-1}R\tr Q
\nonumber\\
&&
+\frac{(n-1)^2}{18}N^nR^2
-\frac{n+3}{90}N^nR_{\mu\nu}R^{\mu\nu}
+\frac{n-2}{90}N^nR_{\mu\nu\alpha\beta}R^{\mu\nu\alpha\beta}.
\eea

Now we compute $h_{1,1}$.
We have
\bea
h_{1,\mu\nu}&=&\bigg[N^n\Phi_{1;(\mu\nu)}
+(N^nJ_1+nN^{n-1}\Lambda_{1})(\Phi_{0;(\mu\nu)}
+2n\zeta_{;\mu\nu})
\nonumber\\
&&
+N^n[J_{1;\m\n}]+nN^{n-1}[\Lambda_{1;\m\n}]
-g_{\mu\nu}h_{2,0}\bigg]\,.
\eea
Therefore,
\bea
h_{1,1}&=&\frac{1}{2n}\bigg[N^n\Phi_1{}^{;\mu}{}_{\mu}
+[N^nJ_1+nN^{n-1}\Lambda_1](\Phi_{0}{}^{;\mu}{}_{\mu}
+2n\zeta^{;\mu}{}_{\mu})
\nonumber\\
&&
+N^n[{{J_{1}}^{;\m}}_\m]+nN^{n-1}[ {{\Lambda_{1}}^{;\m}}_\m]
-nh_{2,0}\bigg]\,.
\eea

We already know $[h_2]$, $[J_1]$, $[\Lambda_1]$, $[\zeta_{;\mu\nu}]$
and $[\Phi_{0;(\mu\nu)}]$.
The only new objects to compute are $\Phi_{1;(\mu\nu)}$, $J_{1;(\m\n)}$, and $\Lambda_{1;(\m\n)}$.
By using the definition of the functions $\chi_{i,j}$ we find that
the coincidence limits of all first derivatives vanish, that is,
\be
[\chi_{i,j,\mu}]=0\,.
\ee
Therefore,
\be
[\Phi_{1;(\mu\nu)}]=\left[-\frac{1}{2}\chi_{1,2;(\mu\nu)}
+\chi_{2,1;(\mu\nu)}
+\chi_{3,1;(\mu\nu)}\right]\,.
\ee
We compute
\bea
[\chi_{1,2;(\mu\nu)}] &=& 4[F_{1,(\mu'\nu')}]
+8[F_{0,\mu'\alpha'}F_0^{\alpha'}{}_{\nu'}],
\\
{}[\chi_{2,1;(\mu\nu)}] &=& -2[E_{1,(\mu';\nu)}]
-2 [E_{0,\alpha';(\mu}F_0^{\alpha'}{}_{\nu')}],
\\
{}[\chi_{3,1;(\mu\nu)}] &=& -2[\tilde E_{1,(\mu';\nu)}]
-2 [F_{0,(\mu' |\alpha'|}\tilde E^{\alpha'}_{0;\nu)}].
\eea
Next, we compute
\bea
[E_{1,\mu';\nu}] &=& -\frac{1}{N}\tr [\nabla_\nu\nabla_{\mu'} a_1]
-\frac{2}{N}\tr [{\cal P}_{;\mu'\nu}]Q\,,
\\
{}[\tilde E_{1,\mu';\nu}] &=& \frac{1}{N}\tr [\nabla_{\nu}\nabla_{\mu} a_1]
+\frac{2}{N}\tr [{\cal P}_{;\mu\nu}]Q\,.
\eea
Therefore,
\bea
[\chi_{1,2;(\mu\nu)}] &=& \frac{4}{N}\tr [\nabla_{(\mu'}\nabla_{\nu)} a_1]
+\frac{4}{9}R_{\mu\alpha}R^\alpha_\nu,
\\
{}[\chi_{2,1;(\mu\nu)}] &=& 
\frac{2}{N}\tr [\nabla_{(\mu'}\nabla_{\nu)} a_1]
+\frac{1}{18}R_{\mu\alpha}R^\alpha_\nu,
\\
{}[\chi_{3,1;(\mu\nu)}] &=& 
-\frac{2}{N}\tr [\nabla_{(\mu}\nabla_{\nu)} a_1]
+\frac{1}{18}R_{\mu\alpha}R^\alpha_\nu,
\eea
and
\be
[\Phi_{1;(\mu\nu)}]=-\frac{2}{N}\tr [\nabla_{(\mu}\nabla_{\nu)} a_1]
-\frac{1}{9}R_{\mu\alpha}R^\alpha_\nu\,.
\ee
Therefore,
\be
[\Phi_{1}{}^{;\mu}{}_{\mu}]=-\frac{2}{N}\tr [\nabla^{\mu}\nabla_{\mu} a_1]
-\frac{1}{9}R_{\mu\nu}R^{\mu\nu}\,.
\ee
By using the known formula for the coefficient 
$[\nabla^{\mu}\nabla_{\mu} a_1]$ (see Appendix) we get
\bea
[\Phi_{1}{}^{;\mu}{}_{\mu}] &=&
-\frac{2}{3N}\tr\nabla^\mu\nabla_{\mu}Q
+\frac{1}{3N}\tr {\cal R}_{\alpha \mu}{\cal R}^{\alpha\mu}
+\frac{2}{15}\nabla^\mu\nabla_\mu R
\\[5pt]
&&
-\frac{2}{15}R_{\mu\nu}R^{\mu\nu}
+\frac{1}{45}R_{\mu\nu\alpha\beta}R^{\mu\nu\alpha\beta}
\,.
\eea

We still need \([J_1{}^{;\m}{}_{\m}]\) and \([\Lambda_1{}^{;\m}{}_\m]\).
\be
J_1 = -2 {{F_0}^{\mu'}}_{\mu'}.
\ee
We have
\be
{{F_0}^{\a'}}_{\b'} = \frac{1}{N} \g^{\a' \d} \tr \left( \zeta_{;\d \b'} 
+ \zeta_{;\d} \zeta_{;\b'} 
+ \zeta_{;\d} {\cal P}^{-1}{\cal P}_{;\b'} 
+ \zeta_{;\b'} {\cal P}^{-1} {\cal P}_{;\d} 
+ {\cal P}^{-1} {\cal P}_{;\d \b'}  \right) \,,
\ee
then
\bea
[{{F_0}^{\a'}}_{\b' ; \mu \nu}] & =&
 - \frac{1}{6} [{\gamma^{\a' \d}}_{;\mu \nu}] R_{\d \b} 
- \frac{1}{N} g^{\a \d} 
\tr \Bigl[ \zeta_{;\d \b' \mu \nu} 
+ 2 \zeta_{;\d (\mu} \zeta_{;|\b'|\nu)} 
+ 2 \zeta_{;\d (\mu} {\cal P}_{;|\b'|\nu)}  
\nonumber\\
&&
+ 2 \zeta_{;\b' ( \mu} {\cal P}_{;|\d | \nu)} 
- {\cal P}_{;\mu \nu} {\cal P}_{;\d \b'} 
+ {\cal P}_{;\d \b' \mu \nu} \Bigr]\,,
\eea
thus
\be
[{{F_0}^{\a'}}_{\b' , \mu \nu}] 
= - \frac{1}{6} [{\gamma^{\a' \d}}_{;\mu \nu}] R_{\d \b} 
-  [ \zeta^{;\a}{}_{\b' \mu \nu}] 
- \frac{2}{36} {R^\a}_{(\mu} R_{|\b|\nu)} 
+ \frac{1}{4N} \tr {\cal R}_{\mu \nu} {{\cal R}^\a }_{\b} 
- \frac{1}{N} \tr [{{\cal P}^{;\a}}_{ \b' \mu \nu} ] .
\ee
Therefore we have
\be
[F_0{}^{\m'}{}_{\m'}{}^{; \n}{}_\n] = - [ \zeta{}^{;\mu}{}_{\mu'}{}^{ \nu}{}_{ \nu}] 
- \frac{1}{N} \tr [  {\cal P}{}^{;\mu}{}_{ \mu'}{}^{ \nu }{}_{\nu} ].
\ee
Simplifying, we find
\be
[F_0{}^{\m'}{}_{\m'}{}^{; \n}{}_\n] = \frac{1}{30} \nabla_\m \nabla^\m R 
+ \frac1{45} R_{\m\n}R^{\m\n} 
+ \frac1{30} R_{\a\b\m\n}R^{\a\b\m\n} 
- \frac{1}{2N} \tr {\cal R}_{\m\n} {\cal R}^{\m\n}.
\ee
Then we have
\be 
[{{J_1}^{;\mu}}_\mu] = 
-2 [ F_0{}^{\mu'}{}_{\mu'}{}^{;\nu}{}_\nu] = 
-\frac{2}{30} \nabla_\m \nabla^\m R 
- \frac2{45} R_{\m\n}R^{\m\n} 
- \frac2{30} R_{\a\b\m\n}R^{\a\b\m\n} 
+ \frac{1}{N} \tr {\cal R}_{\m\n} {\cal R}^{\m\n} .
\ee

Next, calculating \([\Lambda_{1;\mu \nu}]\), we have
\be 
\Lambda_{1;\mu \nu} = 
- \tr ( a_{1;\mu \nu} + a_{1;\mu \nu}^*),
\ee
thus
\be
[ {{\Lambda_{1}}^{;\mu}}_\mu ] = 
-2 \tr [ \nabla_\mu \nabla^\mu a_1].
\ee

Finally, by collecting all the terms, we get
\bea
h_{1,1} &=& 
\frac{n^2-n-2}{6n}N^{n-1}\tr\nabla^\mu\nabla_{\mu}Q
+\frac{2+3n-2n^2}{60n}N^n\nabla^\mu\nabla_\mu R
\nonumber\\
&&
+\frac{8+4n-n^2}{12n}N^{n-1}\tr {\cal R}_{\mu\nu}{\cal R}^{\mu\nu}
-nN^{n-1}\tr Q^2
-n(n-1)N^{n-2}(\tr Q)^2
\nonumber\\
&&
-\frac{n(n+3)}{3}N^{n-1}R\tr Q
+\frac{3n^2+2n-5}{36}N^nR^2
+\frac{n^2+n-16}{180n}N^nR_{\mu\nu}R^{\mu\nu}
\nonumber\\
&&
+\frac{-n^2+4n-4}{180n}N^nR_{\mu\nu\alpha\beta}R^{\mu\nu\alpha\beta}
\,.
\eea


Next, we compute $h_{0,2}$. 
We have
\bea
h_{0,\mu\nu \a \b} &=& 
\bigg[ N^n \Phi_{0} \left( 2n \zeta_{;(\mu \nu \a \b)} 
+ 24 n^2 \zeta_{; (\mu \nu} \zeta_{;\a \b)} \right)
\nonumber\\
&&
+ 6 N^n \zeta_{;(\mu \nu} \Phi_{0;\a \b)} 
- 6 N^{n-1} \zeta_{;(\mu \nu} g_{\a \b)} \left( N J_1 + n \Lambda_1 \right) 
\nonumber\\
&&
+ N^n \Phi_{0;( \mu \nu \a \b )} 
- 6 N^n g_{ (\mu \nu} J_{1; \a\b)} 
- 6nN^{n-1}g_{(\mu \nu} \Lambda_{1; \a \b)} \bigg] \,.
\eea
Therefore,
\bea
h_{0,2} &=& \frac{1}{8n^2} \Bigg( N^n [\Phi_{0}] \bigg( 2n [\zeta_{(4)}] 
+ \frac{ 24 n^2}{108} \left( R^2 + 2  R_{\mu \nu} R^{\a \b} \right) \bigg)
\nonumber\\
&&
+ N^n \frac{1}{3} ( R [{{\Phi_{0;}}^\mu}_\mu] 
+ 2  R^{\mu \nu} [\Phi_{0; (\mu \nu)}] ) 
- N^{n-1} \frac{n+2}{3} R \left( N [J_1] + n [\Lambda_1] \right) 
\nonumber\\
&&
+ N^n g^{\mu \nu} g^{\a \b} [\Phi_{;( \mu \nu \a \b )}] 
- 2 N^n (n+2) [{{J_{1;}}^\mu}_\mu]
\nonumber\\
&&
 - 2N^{n-1}n (n+2) [{{{\Lambda_{1;}}^\mu}_\mu} ] \Bigg) \,.
\eea
We already know \([\zeta_{(4)}]\), \([\Phi_0]\), \([\Phi_{0;(\mu \nu)}]\), \([J_1]\), \([\Lambda_1]\), \([J_1{}^{;\mu}{}_{\mu}]\), and \([\Lambda_1{}^{;\m}{}_{\mu}]\) . We calculate  \( [\Phi_{0; (\a \b \mu \nu)}]\). We find
\be
\Phi_{0; ( \a \b \mu \nu)} = 
-\frac12 \chi_{1,1;(\a \b \mu \nu)} 
+ \chi_{2,0;(\a\b\mu\nu)} 
+ \chi_{3,0;(\a\b\mu \nu)}
+ 6\chi_{2,0;(\a\b|}\chi_{3,0;|\m\n)} 
-6\chi_{1,0;(\a\b|}\chi_{4,0;|\m\n)}.
\ee

We have
\be 
\chi_{1,1;(\a \b \mu \nu)} = 
24 \s_{\d' (\a} F_0{}^{\d' \g'}{}_{;\b\m} \s_{|\g' |\n)} \,,
\ee
then
\be
g^{\a\b}g^{\m\n} [\chi_{1,1;(\a \b \mu \nu)}] = 
8 \left( [F_0{}^{\mu'}{}_{\mu'}{}^{;\nu}{}_\nu ]
+ 2 [F_0{}^{\m' \n'}{}_{;\m\n}] \right)\,.
\ee
We know \([F_0{}^{\mu'}{}_{\mu'}{}^{;\nu}{}_\nu ]\). Next, 
\be
[ F_0{}^{\m'\n'}{}_{;\m\n} ] = -\frac16 [{\g^{\m'\a}}_{;\m\n}]R^{\n}_\a 
- [{\zeta^{;\m\n'}}_{\m\n}] 
+ \frac{1}{36}\left( R^2
+ R_{\m\n}R^{\m\n}\right) 
- \frac{1}{4N} \tr {\cal R}_{\m\n} {\cal R}^{\m\n} 
- \frac{1}{N} [ {{\cal P}^{;\m \nu'}}_{\m\n}].
\ee
Simplifying, we obtain
\be
[F_0{}^{\m'\n'}{}_{;\m\n} ] = 
\frac{1}{36}R^2 
+ \frac{1}{30} \nabla_\m \nabla^\m R 
-\frac{7}{60} R_{\m\n} R^{\m\n} 
+\frac{1}{30} R_{\a\b\m\n}R^{\a\b\m\n}
+ \frac{1}{4N} \tr {\cal R}_{\m\n}{\cal R}^{\m\n}.
\ee
Thus
\be
g^{\a\b}g^{\m\n} [\chi_{1,1;(\a \b \mu \nu)}] = 
\frac{4}{9} R^2 
+ \frac{4}{5} \nabla_\mu \nabla^\m R
- \frac{19}{90} R_{\m\n} R^{\m \n}
+ \frac{1}{10} R_{\a\b\m\n} R^{\a\b\m\n}\,.
\ee

We have
\be
[\chi_{2,0;(\a\b\m\n)}] 
= [4 E_{0,\d' ;(\a\b\m} {\s^{\d'}}_{\n)}]
\ee
So
\be
[E_{0,\d' ;(\a\b\m)}] = 
\frac{1}{N} \tr \left[ \zeta_{;\d' (\a\b\m)} 
+ {\cal P}_{;\d' (\a\b\m)} 
+ 2 {\cal P}^{-1}_{;(\a\b} {\cal P}_{;|\d' |\m)}   \right]
\ee
Thus we find
\be
[\chi_{2,0;(\a\b\m\n)} ]=
 4 [\zeta_{;(\a' \b\m\n)}] 
 + \frac{4}{N} \tr [ {\cal P}_{;(\a'\b\m\n)}] 
 - \frac{12}{N} \tr [ {\cal P}_{;\a\b} {\cal P}_{;\m'\n} ].
\ee
Then
\be
g^{\a\b} g^{\m\n} [\chi_{2,0;(\a\b\m\n)} ] = 
-\frac{2}{15} \nabla_\m \nabla^\m R 
- \frac{2}{135} R_{\m\n }R^{\m\n} 
- \frac{2}{15} R_{\a\b\m\n}R^{\a\b\m\n} 
- \frac{2}{N} \tr {\cal R}_{\m\n} {\cal R}^{\m\n}\,.
\ee

Next, we have
\be
[\chi_{3,0;(\a\b\m\n)} ]= [4 \s_{\d' (\a}   \tilde{E}_0^{\d'}{} _{;\b\m\n)}].
\ee
We compute
\bea
[\tilde E^{\a'}_{0;(\b\m\n)} ] &=& [\g^{\a\d}] \left( [\zeta_{;\d (\b\m\n)}] 
+ \frac{1}{N} \tr [{\cal P}_{;\d' (\b\m\n)}] 
+ \frac{3}{N} \tr [{\cal P}^{-1}_{;(\b\m} {\cal P}_{;|\d|\n)}]   \right)
\nonumber\\
&& 
+ \frac{1}{N} [{\g^{\a' \d}}_{;(\b\m}] \tr [\zeta_{;|\d | \n)}] 
+ \frac{1}{N}[{\g^{\a' \d}}_{;(\b\m}] \tr [{\cal P}_{;|\d | \n)}].
\eea
Then
\be
[\chi_{3,0;(\a\b\m\n)}] = 
4 [\zeta_{;(\a\b\m\n)} ]
- \frac{12}{N} \tr [ {\cal P}_{;(\a\b} {\cal P}_{;\m\n)}]
-4 [{\g^{\d' \g}}_{;(\a\b}] [\zeta_{;|\g| \m}] g_{|\d| \n)}\,.
\ee
Then we obtain
\be
g^{\a\b}g^{\m\n} [\chi_{3,0;(\a\b\m\n)}] = 
\frac{4}{5} \nabla_\mu \nabla^\mu R 
+ \frac{4}{45} R_{\mu \nu} R^{\mu \nu}
+ \frac{2}{15} R_{\a\b\m\n}R^{\a\b\m\n}
-\frac{2}{N} \tr {\cal R}_{\m\n} {\cal R}^{\m\n} \,.
\ee

Lastly we need
\be
[\chi_{4,0;{\m\n}}] = [ E_{0,\a'(\m} \tilde E^{\a'}_{0,\n)}]\,.
\ee
We find
\be 
[E_{0,\a'\m}] = - \frac{1}{6} R_{\a\m}
\ee
and
\be
[\tilde E^{\a'}_{0,\n)}] = - \frac{1}{6} {R^{\a}}_{\m},
\ee
thus
\be
[\chi_{4,0;{(\m\n)}}] = \frac{1}{36} R_{\a(\m} {R^\a}_{\nu)}\,.
\ee

Thus we have
\be
g^{\a\b}g^{\m\n}[\chi_{2,0;(\a\b}\chi_{3,0;\m\n)}] = \frac{1}{27} ( R^2 + 2 R_{\m\nu} R^{\m\n})
\ee
and
\be
g^{\a\b}g^{\m\n}[\chi_{1,0;(\a\b}\chi_{4,0;\m\n)}] = \frac{n+2}{54} R_{\m\nu} R^{\m\n}.
\ee

Putting this all together, we obtain
\be
g^{\a\b}g^{\m\n}[ \Phi_{0;(\a\b\m\n)}] = \frac{4}{15} \nabla_\m \nabla^\m R 
- \frac{60n-217}{540} R_{\m\n}R^{\m\n} 
- \frac{1}{20} R_{\a\b\m\n}R^{\a\b\m\n}
- \frac{4}{N} \tr {\cal R}_{\m\n} {\cal R}^{\m\n},
\ee
which gives
\bea
h_{0,2} &=& \frac{n^2-n+3}{72n^2} N^n R^2 
+ \frac{n^2+8}{120n^2} N^n \nabla_\m \nabla^\m R
+ \frac{264n^2 + 60n + 433}{4320n^2} N^n R_{\m\n} R^{\m\n}
\nonumber\\
&&
- \frac{8n^2-20n-39}{1440n^2} N^n R_{\a\b\m\n} R^{\a\b\m\n}
- \frac{2n^2+10n+24}{24n^2} N^{n-1} \tr {\cal R}_{\m\n} {\cal R}^{\m\n}
\nonumber\\
&&
+ \frac{n+2}{12n} N^{n-1} R \tr Q
+ \frac{n+2}{6n} N^{n-1} \nabla_\m \nabla^\m \tr Q.
\eea

Thus we finally have
\bea
b_2 &=&
N^n\Biggl\{
\frac{20n^4-8n^3-11n^2-6n+6}{144n^2} R^2
+ \frac{4n^3+11n^2+n-4}{120n^2} \nabla_\m \nabla^\m R
\nonumber\\
&&
+ \frac{-24n^3+84n^2-576n+385}{4320n^2} R_{\m\n}R^{\m\n}
+ \frac{8n^3-8n^2-18n+15}{1440n^2} R_{\a\b\m\n}R^{\a\b\m\n}
\Biggr\}
\nonumber\\
&&
+ \frac{n^3-n^2+3n-12}{12n^2} N^{n-1} \tr {\cal R}_{\m\n} {\cal R}^{\m\n}
+ \frac{-12n^3-4n^2+n+2}{12n} N^{n-1} R \tr Q
\nonumber\\
&&
- \frac{n+2}{6} N^{n-1} \tr \nabla^\m \nabla_\m Q
+ n N^{n-1} \tr Q^2
+ n(n-1) N^{n-2} ( \tr Q )^2 \,.
\eea

\appendix
\section{Appendix}
\setcounter{equation}{0}

\subsection{Derivatives of Synge Function}

We list the coincidence limits of the symmetrized derivatives of the
two-point functions introduced in Sec. 5.
A very efficient algorithm for computing such coincidence limits of the
derivatives of two-point functions is developed in \cite{avramidi91b,avramidi00}.
All the formulas below are computed there.

The coincidence limits of
mixed derivatives of a two-point 
function $f=f(x,x')$ can be computed by
using the equation
\bea
[\nabla_{\mu'}f] &=& \nabla_\mu[f]-[\nabla_{\mu}f].
\eea

First, the coincidence limits of the Synge function
and its first derivatives vanish,
\be
[\sigma]=[\sigma^\mu]=[\sigma^{\mu'}]=0\,.
\ee
The coincidence limits of the second derivatives are
\be
[\sigma_{\mu\nu}]=-[\sigma_{\mu\nu'}]=g_{\mu\nu}\,,
\ee
and the coincidence limits of higher-order symmetrized
derivatives of the vectors $\sigma^\mu$ and $\sigma^{\mu'}$ vanish,
\be
[\sigma^\alpha{}_{(\mu_1\dots\mu_k)}]
=[\sigma^{\alpha'}{}_{(\mu_1\dots\mu_k)}]=0,
\qquad k\ge 2\,.
\ee
All other coincidence limits can be obtained from these by
commuting covariant derivatives.


The coincidence 
limits of the low-order symmetrized derivatives have the form
\cite{avramidi00,avramidi13}
\bea
{}[\sigma^\alpha{}_{\beta\mu}]
&=& {}[\sigma^{\alpha'}{}_{\beta\mu}]=0\;,
\\
{}[\sigma^\alpha{}_{\beta\mu_1\mu_2}]
&=&-\frac{2}{3}R^\alpha{}_{(\mu_1\vert\beta\vert\mu_2)}\;,
\\
{}[\sigma^{\alpha'}{}_{\beta (\mu_1\mu_2)}]
&=& -\frac{1}{3}R^\alpha{}_{(\mu_1|\beta|\mu_2)}\;,
\\
{}[\sigma^{\alpha'}{}_{\beta \mu_1\mu_2}]
&=& -\frac{2}{3}R^\alpha{}_{(\beta\mu_1)\mu_2}\;,
\\
{}[\sigma^{\alpha'}{}_{\beta'\mu_1\mu_2}]
&=& -\frac{2}{3}R^\alpha{}_{(\mu_1|\beta|\mu_2)},
\\
{}[\sigma^\alpha{}_{\beta(\mu_1\mu_2\mu_3)}]
&=& -\frac{3}{2}\nabla_{(\mu_1}R^\alpha{}_{\mu_2\vert\beta\vert\mu_3)}\;,
\\
{}[\sigma^{\alpha'}{}_{\beta(\mu_1\mu_2\mu_3)}]
&=& -\frac{1}{2}\nabla_{(\mu_1}R^\alpha{}_{\mu_2\vert\beta\vert\mu_3)}\;,
\\
{}[\sigma^\alpha{}_{\beta(\mu_1\mu_2\mu_3\mu_4)}]
&=&-\frac{12}{5}\nabla_{(\mu_1}\nabla_{\mu_2}
R^\alpha{}_{\mu_3\vert\beta\vert\mu_4)} 
-\frac{8}{15}R^\alpha{}_{(\mu_1\vert\gamma\vert\mu_2} 
R^\gamma{}_{\mu_3\vert\beta\vert\mu_4)}\;,
\\
{}[\sigma^{\alpha'}{}_{\beta(\mu_1\mu_2\mu_3\mu_4)}]
&=&-\frac{3}{5}\nabla_{(\mu_1}\nabla_{\mu_2}
R^\alpha{}_{\mu_3\vert\beta\vert\mu_4)} 
-\frac{7}{15}R^\alpha{}_{(\mu_1\vert\gamma\vert\mu_2} 
R^\gamma{}_{\mu_3\vert\beta\vert\mu_4)},
\eea

\bea
[\gamma^\alpha{}_{\beta'}]
&=& -\delta^\alpha_\beta,
\\
{}[\gamma^\alpha{}_{\beta';\mu}]
&=& 0,
\\
{}[\gamma^\alpha{}_{\beta';(\mu_1\mu_2)}]
&=&  \frac{1}{3}R^\alpha{}_{(\mu_1\vert\beta\vert\mu_2)}\;,
\\
{}[\gamma^\alpha{}_{\beta';\mu_1\mu_2}]
&=&\frac{2}{3}R^\alpha{}_{(\beta\mu_1)\mu_2}
\;,
\\
{}[\gamma^\alpha{}_{\beta';(\mu_1\mu_2\mu_3)}]
&=& \frac{1}{2}\nabla_{(\mu_1}R^\alpha{}_{\mu_2\vert\beta\vert\mu_3)}\;,
\\
{}[\gamma^\alpha{}_{\beta';(\mu_1\mu_2\mu_3\mu_4)}]
&=&\frac{3}{5}\nabla_{(\mu_1}\nabla_{\mu_2}
R^\alpha{}_{\mu_3\vert\beta\vert\mu_4)} 
-\frac{1}{5}R^\alpha{}_{(\mu_1\vert\gamma\vert\mu_2} 
R^\gamma{}_{\mu_3\vert\beta\vert\mu_4)}\;.  
\eea

The coincidence limit of the
function $\zeta=\frac{1}{2}\log\Delta$ and its derivative vanish
\be
[\zeta]=[\zeta_{;\mu}]=0.
\ee
By using the eq. (\ref{zeta0}) we find 
for the higher-order derivatives
\be
[\zeta_{;(\mu_1\dots\mu_k)}]=-\frac{1}{2}
[\sigma^\alpha{}_{\alpha (\mu_1\dots\mu_k)}],
\qquad k\ge 1\,.
\ee
All higher-order derivatives can be computed by commuting derivatives.

We will need the following derivatives
\bea
{}[\zeta_{;\mu_1\mu_2}]
&=& \frac{1}{6}R_{\mu_1\mu_2}\;,
\\
{}[\zeta_{;\mu\nu'}]
&=& -\frac{1}{6}R_{\mu\nu}\;,
\\
{}[\zeta_{;(\mu_1\mu_2\mu_3)}]
&=& \frac{1}{4}\nabla_{(\mu_1}R_{\mu_2\mu_3)}\;,
\\
{}[\zeta_{;(\mu_1\mu_2\mu_3\mu_4)}]
&=& \frac{3}{10}\nabla_{(\mu_1}\nabla_{\mu_2}R_{\mu_3\mu_4)}
 + \frac{1}{15}R_{\alpha (\mu_1\vert\gamma\vert\mu_2}
R^{\gamma}{}_{\mu_3}{}^{\alpha}{}_{\mu_4)}\;.
\eea

We compute some contractions
\bea
g^{\mu_1\mu_2}g^{\mu_3\mu_4}
[\zeta_{;(\mu_1\mu_2\mu_3\mu_4)}]
&=&\frac{1}{5}\nabla_\mu\nabla^\mu R
+\frac{1}{45}R_{\mu\nu}R^{\mu\nu}
+\frac{1}{30}R_{\mu\nu\alpha\beta}R^{\mu\nu\alpha\beta},
\\
{}[\zeta^{;\m'}{}_\m{}^\n{}_\n ] &=&
-\frac{1}{30}\nabla_\mu\nabla^\mu R
-\frac{1}{45}R_{\mu\nu}R^{\mu\nu}
-\frac{1}{30}R_{\mu\nu\alpha\beta}R^{\mu\nu\alpha\beta},
\\
{}[\zeta^{;\m\n'}{}_{\m\n} ] &=&
-\frac{1}{30}\nabla_\mu\nabla^\mu R
+\frac{4}{45}R_{\mu\nu}R^{\mu\nu}
-\frac{1}{30}R_{\mu\nu\alpha\beta}R^{\mu\nu\alpha\beta},
\eea

For the operator of parallel transport, the coincidence limit
is equal to the identity matrix,
\be
[{\cal P}]=\II
\ee
and the coincidence limits of higher-order symmetrized
derivatives vanish, 
\be
[{\cal P}_{;(\alpha\mu_1\dots\mu_k)}]
=[{\cal P}_{;(\alpha'\mu_1\dots\mu_{k})}]
=0\,,\qquad k\ge 0.
\ee
In particular, it is easy to get
\be
{}[{\cal P}_{;\mu}] = {}[{\cal P}_{;\mu'}] = 0\,.
\ee
All other coincidence limits can be obtained from these
by commuting derivatives.

It is convenient to introduce the following vector \cite{avramidi00,avramidi91b}
\be
\mathcal{A}_{\mu'}={\cal P}^{-1}\gamma^\nu{}_{\mu'}\nabla_\nu{\cal P},
\ee
so that the first derivative of the parallel transport operator has the form
\be
{\cal P}_{;\nu}={\cal P}\sigma^{\mu'}{}_{\nu}{\cal A}_{\mu'} \,.
\ee
It is easy to see that
\be
[{\cal A}_{\mu'}] =0\,.
\ee
Further, some low-order symmetrized derivatives of the vector ${\cal A}_{\mu'}$ are
\bea
[{\cal A}^{\nu'}{}_{;\mu_1}]  
&=&  \frac{1}{2}{\cal R}^\nu{}_{\mu_1}\;,  
\\{}
[{\cal A}^{\nu'}{}_{;(\mu_1\mu_2)}]  
&=&  \frac{2}{3}\nabla_{(\mu_1}{\cal R}^\nu{}_{\mu_2)}\;,
\\{}
[{\cal A}^{\nu'}{}_{;(\mu_1\mu_2\mu_3)}]
&=& \frac{3}{4}\nabla_{(\mu_1}\nabla_{\mu_2}{\cal R}^{\nu}{}_{\mu_3)}
-\frac{1}{4}R^\nu{}_{(\mu_1|\alpha|\mu_2}{\cal R}^\alpha{}_{\mu_3)}\;.
\eea

By using the covariant Taylor expansion of this vector one can compute all 
coincidence limits of the parallel transport operator we need.
Some low-order derivatives have the form
\bea
{}[{\cal P}_{;\nu\mu}] &=& -[{\cal P}_{;\nu\mu'}]=-\frac{1}{2}{\cal R}_{\nu\mu},
\\
{}[{\cal P}^{;\nu}{}_{\mu_1\mu_2}]&=&
-\frac{2}{3}\nabla_{(\mu_1}\mathcal{R}^\nu{}_{\mu_2)},
\\
{}[{\cal P}^{;\nu'}{}_{\mu_1\mu_2}]&=&
-\frac{1}{3}\nabla_{(\mu_1}\mathcal{R}^{\nu}{}_{\mu_2)},
\\
{}[{\cal P}^{;\nu}{}_{(\mu_1\mu_2\mu_3)}]&=&
-\frac{3}{4}\nabla_{(\mu_1}\nabla_{\mu_2}\mathcal{R}^\nu{}_{\mu_3)}
-\frac{1}{4}R^\nu{}_{(\mu_1|\alpha|\mu_2}{\cal R}^\alpha{}_{\mu_3)}\;,
\\
{}[{\cal P}^{;\nu'}{}_{(\mu_1\mu_2\mu_3)}]&=&
\frac{1}{12}\nabla_{(\mu_1}\nabla_{\mu_2}\mathcal{R}^\nu{}_{\mu_3)}
+\frac{1}{4}R^\nu{}_{(\mu_1|\alpha|\mu_2}{\cal R}^\alpha{}_{\mu_3)}\;.
\eea
We will need the following contractions
\bea
{}\tr [{\cal P}^{;\mu}{}_{\mu'}{}^{\nu}{}_{\nu}]&=&
\frac{1}{2} {\cal R}_{\m\n} {\cal R}^{\m\n} \;,
\\
{}\tr [{\cal P}^{;\mu \nu'}{}_{\mu \nu}]&=&
-\frac{1}{2} {\cal R}_{\m\n} {\cal R}^{\m\n} \;,
\\
{}\tr [{\cal P}^{;\mu \nu'}{}_{\nu \mu}]&=&
0 \;.
\eea

\subsection{Derivatives of the Heat Kernel Coefficients}

We list below the coincidence limits of the heat kernel coefficients
and their derivatives
(see \cite{avramidi00,avramidi13}; notice some different
sign conventions)
\bea
[a_0]&=&\II\,,
\\
{}[a_1]&=&Q-\frac{1}{6}R\,,
\label{b1}
\\[5pt]
[\nabla_\mu a_1] &=& 
\frac{1}{2}\nabla_\mu Q
-\frac{1}{12}\nabla_\mu R
+\frac{1}{6}\nabla_\nu{\cal R}^\nu{}_\mu\,,
\\[5pt]
{}[\nabla_{(\mu}\nabla_{\nu)}a_1] &=&
\frac{1}{3}\nabla_{(\mu}\nabla_{\nu)}Q
-\frac{1}{6}{\cal R}_{\alpha(\mu}{\cal R}^{\alpha}{}_{\nu)}
+\frac{1}{6}\nabla_{(\mu}\nabla_{|\alpha|}{\cal R}^\alpha{}_{\nu)}
-\frac{1}{20}\nabla_\mu\nabla_\nu R
\\[5pt]
&&
-\frac{1}{60}\nabla_\alpha\nabla^\alpha R_{\mu\nu}
+\frac{1}{45}R_{\mu\alpha}R^\alpha{}_{\nu}
-\frac{1}{90}R_{\mu\alpha\beta\gamma}R_\nu{}^{\alpha\beta\gamma}
-\frac{1}{90}R_{\alpha\beta}R^\alpha{}_\mu{}^\beta{}_\nu{}
\,,
\nonumber\\[5pt]
{}[\nabla^\mu\nabla_{\mu}a_1] &=&
\frac{1}{3}\nabla^\mu\nabla_{\mu}Q
-\frac{1}{6}{\cal R}_{\alpha \mu}{\cal R}^{\alpha\mu}
-\frac{1}{15}\nabla^\mu\nabla_\mu R
\\[5pt]
&&
+\frac{1}{90}R_{\mu\nu}R^{\mu\nu}
-\frac{1}{90}R_{\mu\nu\alpha\beta}R^{\mu\nu\alpha\beta}
\,,
\nonumber\\[5pt]
{}[a_2] &=&\left(Q-\frac{1}{6}R\right)^2
-\frac{1}{3}\nabla^\mu\nabla_\mu Q
+\frac{1}{6}{\cal R}_{\mu\nu}{\cal R}^{\mu\nu}
+\frac{1}{15}\nabla^\mu\nabla_\mu R
\nonumber\\[5pt]
&&-\frac{1}{90} R_{\mu\nu}R^{\mu\nu}
+\frac{1}{90}R_{\mu\nu\alpha\beta}R^{\mu\nu\alpha\beta}\,.
\label{b2}
\eea

\subsection{Gaussian Integrals}

Let $A=(A_{ij})$ be a real symmetric positive 
matrix. 
Then for any vector $B=(B_i)$ there holds
\bea
&&
\int\limits_{\RR^n} d\xi\;
\exp\Bigl(-\left<\xi,A\xi\right>
+\left<B,\xi\right>\Bigr)
\nonumber\\
&&\qquad
=\pi^{n/2}(\det A)^{-1/2}
\exp\left(\frac{1}{ 4}\,\left<B,A^{-1}B\right>\right)\,,
\label{218cc}
\eea
where $A^{-1}=(A^{ij})$ is the inverse of the matrix $A$. 
By expanding both sides of eq. (\ref{218cc}) in Taylor series in 
$B_i$ we obtain
\bea
&&
\int\limits_{\RR^n}d\xi\;
\exp\Bigl(-\left<\xi,A\xi\right>\Bigr)\, \xi^{i_1}\cdots \xi^{i_{2k+1}}=0\,,
\\[10pt]
&&
\int\limits_{\RR^n}d\xi\;
\exp\Bigl(-\left<\xi,A\xi\right>\Bigr)\,
\xi^{i_1}\cdots \xi^{i_{2k}}
\nonumber\\
&&\qquad
=\pi^{n/2}(\det A)^{-1/2}\,\frac{(2k)!}{ 2^{2k}k!}\,
A^{(i_1\,i_2}\cdots A^{i_{2k-1}\,i_{2k})}\,,
\label{gauss1}
\eea
where the parenthesis
denote complete symmetrization over all indices included.

We introduce the Gaussian average
\be
\left<f\right>
=\pi^{-n/2}(\det A)^{1/2}\int\limits_{\RR^n}d\xi\;
\exp\Bigl(-\left<\xi,A\xi\right>\Bigr)f(\xi)\,.
\ee
Then, the above formulas can be written in the form
\bea
\left<\xi^{i_1}\cdots \xi^{i_{2k+1}}\right>&=&0\,,
\\[10pt]
\left<\xi^{i_1}\cdots \xi^{i_{2k}}\right>
&=&\frac{(2k)!}{ 2^{2k}k!}\,
A^{(i_1\,i_2}\cdots A^{i_{2k-1}\,i_{2k})}\,.
\label{gauss2}
\eea

\subsection{Leibnitz Rule}

In this paper we extensively use the Leibnitz rule for the
symmetrized derivative of the product
\be
\nabla_{(\mu_1}\cdots\nabla_{\mu_n)}(fg)
=\sum_{k=0}^n {n\choose k} \left(\nabla_{(\mu_1}\cdots\nabla_{\mu_k}f\right)
\left(\nabla_{\mu_{k+1}}\cdots\nabla_{\mu_n)}g\right).
\ee
In particular,
\bea
\nabla_{(\mu_1}\cdots\nabla_{\mu_4)}(fg)
&=&f_{;(\mu_1\mu_2\mu_3\mu_4)}g
+4f_{;(\mu_1\mu_2\mu_3}g_{;\mu_4)}
+6f_{;(\mu_1\mu_2}g_{;\mu_3\mu_4)}
\nonumber\\
&&
+4f_{;(\mu_1}g_{;\mu_2\mu_3\mu_4)}
+fg_{;(\mu_1\mu_2\mu_3\mu_4)}.
\eea

Let $f=e^h$. Then
\bea
f_{;\mu}&=& fh_{;\mu},\\
f_{;(\mu_1\mu_2)}&=&f\left(h_{;(\mu_1\mu_2)}
+h_{;(\mu_1}h_{;\mu_2)}\right),\\
f_{;(\mu_1\mu_2\mu_3)}&=& f\left(h_{;(\mu_1\mu_2\mu_3)}
+3h_{;(\mu_1\mu_2}h_{;\mu_3)}
+h_{;(\mu_1}h_{;\mu_2}h_{;\mu_3)}\right),\\
f_{;(\mu_1\mu_2\mu_3\mu_4)}
&=&f\Biggl(h_{;(\mu_1\mu_2\mu_3\mu_4)}
+3h_{;(\mu_1\mu_2}h_{;\mu_3\mu_4)}
+4h_{;(\mu_1}h_{;\mu_2\mu_3\mu_4)}
\nonumber\\
&&
+6h_{;(\mu_1}h_{;\mu_2}h_{;\mu_3\mu_4)}
+h_{;(\mu_1}h_{;\mu_2}h_{;\mu_3}h_{;\mu_4)}
\Biggr).
\eea



\begin{thebibliography}{99}


\bibitem{avramidi91b} I. G. Avramidi, {\it A covariant technique for the
calculation of the one-loop effective action}, Nuclear Phys. B \textbf{355}
(1991) 712--754; Erratum: Nuclear Phys. B {\bf 509} (1998) 557--558

\bibitem{avramidi00} I. G. Avramidi, \textit{Heat Kernel and Quantum Gravity},
Berlin: Springer, 2000

\bibitem{avramidi02} I. G. Avramidi, {\it Heat kernel approach in quantum field
theory}, Nuclear Phys. B Proc. Suppl. {\bf 104} (2002) 3--32

\bibitem{avramidi10} I. G. Avramidi, {\it Mathemathical tools for calculation
of the effective action in quantum gravity}, in: {New Paths Towards Quantum
Gravity}, Eds. B. Booss-Bavnbek, G. Esposito and M. Lesch, Berlin, Springer,
2010, pp. 193-259

\bibitem{avramidi13} I. G. Avramidi, {\it Heat Kernel: with Applications to
Finance}, 360 pp. (World Scientific, 2014). Under review

\bibitem{avramidi09c} I. G. Avramidi and G. Fucci, {\it Non-perturbative heat
kernel asymptotics on homogeneous Abelian bundles}, {Comm. Math. Phys.} {\bf
291} (2009) 543-577

\bibitem{berger03} M. Berger, \textit{A Panoramic View of Riemannian Geometry},
Berlin: Springer, 1992

\bibitem{berline92} N. Berline, E. Getzler and M. Vergne, \textit{Heat Kernels
and Dirac Operators}, Berlin: Springer, 1992

\bibitem{branson95} T. Branson, {\it Sharp inequalities, the functional
determinant, and the complementary series}, Trans. Amer. Math. Soc. {\bf 347}
(1995) 3671--3742


\bibitem{branson92b} T. Branson, S.-Y.A. Chang and P. Yang, {\it Estimates and
extremals for zeta function determinants on four-manifolds}, Comm. Math.
Phys. {\bf 149} (1992) 241--262
     

\bibitem{dewitt65} B.S. De~Witt, {\it Dynamical Theory of Groups and Fields},
New York: Gordon and Breach, 1965



\bibitem{egorov98} Yu. V. Egorov and M. A. Shubin, {\it Foundations of the
Classical Theory of Partial Differential Equations}, Berlin, Springer,
1998

\bibitem{gilkey75b} P. B. Gilkey, {\it The spectral geometry of Riemannian
manifold}, J. Differential Geom. {\bf 10} (1975) 601--618.

\bibitem{gilkey95} P. B. Gilkey, \textit{Invariance Theory, the Heat Equation
and the Atiyah--Singer Index Theorem}, Boca Raton: CRC Press, 1995

\bibitem{grigoryan09}
A. Grigor'yan, {\it Heat Kernel and Analysis on Manifolds},
AMS, International Press, 2009

\bibitem{hadamard23} J. Hadamard, \textit{Lectures on Cauchy's Problem}, in:
{\it Linear Partial Differential Equations}, New Haven: Yale University Press,
1923



\bibitem{hurt83} N. E. Hurt, {\it Geometric Quantization in Action:
Applications of Harmonic Analysis in Quantum Statistical Mechanics and Quantum
Field Theory}, Berlin: Springer, 1983

\bibitem{kirsten01} K. Kirsten, {\it Spectral Functions in Mathematics and
Physics}, Boca Raton: CRC Press, 2001

\bibitem{kubawara82} R. Kubawara, {\it On isospectral deformations of
Riemannian metrics II}, Compos. Math. {\bf 47} (1982) 195--205

\bibitem{synge60} J.L. Synge, {\it Relativity: The General Theory}, Amsterdam:
North-Holland, 1960

\bibitem{vandeven98} A. E. M. van de Ven, {\it Index free heat kernel
coefficients}, Class. Quant. Grav. {\bf 15} (1998) 2311--2344

\bibitem{vassilevich03} D. V. Vassilevich, {\it Heat kernel expansion: user's
manual}, Phys. Rep. {\bf 388} (2003) 279-360



\end{thebibliography}
\end{document}